\documentclass[12pt]{iopart}

\usepackage{graphicx}
\usepackage{amssymb,amsmath}
\usepackage{bm}
\usepackage{float}
\usepackage{color}
\usepackage{url}
\usepackage{hyperref}
\usepackage{animate}
\usepackage{array}
\usepackage{multirow}
\usepackage{booktabs}
\usepackage{rotating}
\usepackage{lscape}
\newcommand{\labelphantom}[1]{%
  \phantom{\label{#1}}%
}

\begin{document}

\title[Exploring the fusion power plant design space]{Exploring the fusion power plant design space: comparative analysis of positive and negative triangularity tokamaks through optimization}

\author{T Slendebroek$^{1,2}$, A O Nelson$^3$, O M Meneghini$^2$, G Dose$^2$, 
A G Ghiozzi$^2$, J Harvey$^2$, B C Lyons$^2$, J McClenaghan$^2$, 
T F Neiser$^2$, D B Weisberg$^2$, M G Yoo$^2$, E Bursch$^3$, C Holland$^1$}

\address{$^1$ Center for Energy Research, University of California, San Diego, La Jolla, CA 92093-0417, USA}
\address{$^2$ General Atomics, San Diego, California 92121, USA}
\address{$^3$ Columbia University, New York, NY 10027, USA}

\ead{tslendebroek@ucsd.edu}

\begin{abstract}
The optimal configuration choice between positive triangularity (PT) and negative triangularity (NT) tokamaks for fusion power plants hinges on navigating different operational constraints rather than achieving specific plasma performance metrics. This study presents a systematic comparison using constrained multi-objective optimization with the integrated FUsion Synthesis Engine (FUSE) framework. Over 200,000 integrated design evaluations were performed exploring the trade-offs between capital cost minimization and operational reliability (maximizing $q_{95}$) while satisfying engineering constraints including 250 $\pm$ 50 MW net electric power, tritium breeding ratio $>$1.1, power exhaust limits and an hour flattop time. Both configurations achieve similar cost-performance Pareto fronts through contrasting design philosophies. PT, while demonstrating resilience to pedestal degradation (compensating for up to 40\% reduction), are constrained to larger machines ($R_0$ $>$ 6.5 m) by the narrow operational window between L-H threshold requirements and the research-established power exhaust limit ($P_{sol}/R$ $<$ 15 MW/m). This forces optimization through comparatively reduced magnetic field ($\sim$8T). NT configurations exploit their freedom from these constraints to access compact, high-field designs ($R_0 \sim 5.5$ m, $B_0$ $>$ 12 T), creating natural synergy with advancing HTS technology. Sensitivity analyses reveal that PT's economic viability depends critically on uncertainties in L-H threshold scaling and power handling limits. Notably, a 50\% variation in either could eliminate viable designs or enable access to the compact design space. These results suggest configuration selection should be risk-informed: PT offers the lowest-cost path when operational constraints can be confidently predicted, while NT is robust to large variations in constraints and physics uncertainties.
\end{abstract}

\pacs{52.55.Fa, 52.55.-s, 28.52.-s}

\maketitle

\section{Introduction}

Fusion power plant design presents a complex multi-disciplinary optimization challenge that spans plasma physics, engineering systems, and economic constraints. Contemporary design methodologies typically use low-fidelity analytical models to survey broad parameter spaces and identify promising design points based on specific technological advantages, such as high magnetic field strength, spherical tokamak geometry, advanced plasma shaping, or novel physics regimes. These preliminary designs are subsequently refined through detailed analyses by specialized expert teams with domain-specific knowledge. However, this approach inherently encounters fundamental integration challenges: expert analyses often reveal inconsistencies with initial design assumptions, leading to iterative refinements that cannot be efficiently incorporated back into the original low-fidelity framework. This results in labor-intensive manual integration of disconnected analyses, limiting both computational efficiency and the scope of design space exploration.

To address these methodological limitations, we have developed the FUsion Synthesis Engine (FUSE\cite{meneghini2024fuse}): an integrated computational framework that consolidates expert knowledge within a unified modeling environment. Rather than treating specialized analyses as external processes, FUSE incorporates domain expertise directly into a centralized data structure, enabling interaction between all design aspects. For instance, the framework seamlessly integrates plasma equilibrium and transport calculations that determine fusion power, which then drive neutronics analysis for breeding blanket design and first wall heat loads. These plasma solutions simultaneously inform engineering subsystems ranging from magnet stress analysis to cooling system requirements, while economic models evaluate capital costs based on the complete integrated design. This tight coupling ensures that design modifications propagate consistently throughout the facility model, capturing inter-dependencies that traditional sequential approaches miss.

The comprehensive nature of FUSE enables systematic design space exploration through constrained multi-objective optimization (CMOOP). This methodology allows precise formulation of design objectives and constraints, with automated computational search replacing manual parameter surveys. We use genetic algorithms to identify Pareto-optimal solution families that satisfy engineering constraints, such as net electric power output exceeding 200 MW, while optimizing competing objectives including capital cost minimization and minimizing operational risk. Although genetic algorithms may not converge as rapidly as gradient-based methods, their exceptional parallelizability enables evaluation of hundreds of thousands of integrated designs per day on moderate computational clusters, facilitating comprehensive design space studies.

To demonstrate FUSE's capabilities for comparative power plant design studies, we apply this multi-objective optimization framework to evaluate Negative Triangularity (NT) tokamak concepts against conventional Positive Triangularity (PT) configurations for fusion power plants. NT plasmas, characterized by triangularity $\delta < 0$, have emerged as promising power plant candidates due to their intrinsic ELM-free operation, enhanced power handling characteristics, and improved core performance relative to PT L-mode plasmas \cite{Camenen2005, pochelon_energy_1999, Austin2019, marinoni_brief_2021, nelson_robust_2023, thome_overview_2024, scotti_high_2024, nelson_first_2024, aucone_experiments_2024, paz-soldan_simultaneous_2024, nelson_prospects_2022}. While preliminary NT power plant designs have been investigated \cite{Kikuchi2019, Medvedev2015, the_manta_collaboration_manta_2024, schwartz_dee_2022, wilson_characterizing_2024, miller_power_2024}, the configuration remains significantly under-explored compared to conventional PT approaches, primarily due to its recent experimental exploration. The FUSE framework enables systematic exploration of the NT design space with realistic high-fidelity physics models and comprehensive engineering constraints, facilitating quantitative assessment of NT advantages and limitations relative to PT configurations for fusion power plant applications.

This comparative optimization study reveals that despite different physics constraints and operational characteristics, both NT and PT configurations can achieve similar cost-performance trade-offs, though through markedly different design strategies. PT designs are constrained to larger machine sizes $R_0 > 6.5$m by power exhaust limitations, optimizing costs through reduced magnetic field strength $\sim 8$T. Conversely, NT configurations access more compact, high-field design spaces $R_0 \sim 5.5$ m, $B_0 > 12$ T enabled by relaxed power exhaust constraints, achieving cost parity through different technological approaches. These findings challenge conventional assumptions about configuration-dependent power plant economics and highlight the value of systematic multi-objective optimization for fusion power plant design.

The remainder of this paper is organized as follows. Section 2 describes the FUSE modeling framework, emphasizing the distinct edge physics models required for PT and NT configurations, along with comprehensive experimental validation. Section 3 details the multi-objective optimization formulation, including actuators, constraints, and the genetic algorithm implementation. Section 4 presents comprehensive optimization results, analyzing Pareto fronts, power balance characteristics, and sensitivity to physics uncertainties. Finally, Section 5 discusses the implications for power plant design strategy and outlines future research directions.

\section{Fusion Power Plant Optimization with FUSE}
\label{sec:fuse_methodology}
The comparative analysis of NT and PT configurations requires a modeling framework capable of capturing their different physics while maintaining consistent engineering and economic constraints. This section describes how we extend the FUSE framework \cite{meneghini2024fuse} to enable fair comparison between these configurations, with particular emphasis on the development of a new edge model for NT plasmas.
\subsection{Integrated modeling approach}

\begin{figure*}[ht]
    \centering
    \includegraphics[width=0.99\textwidth]{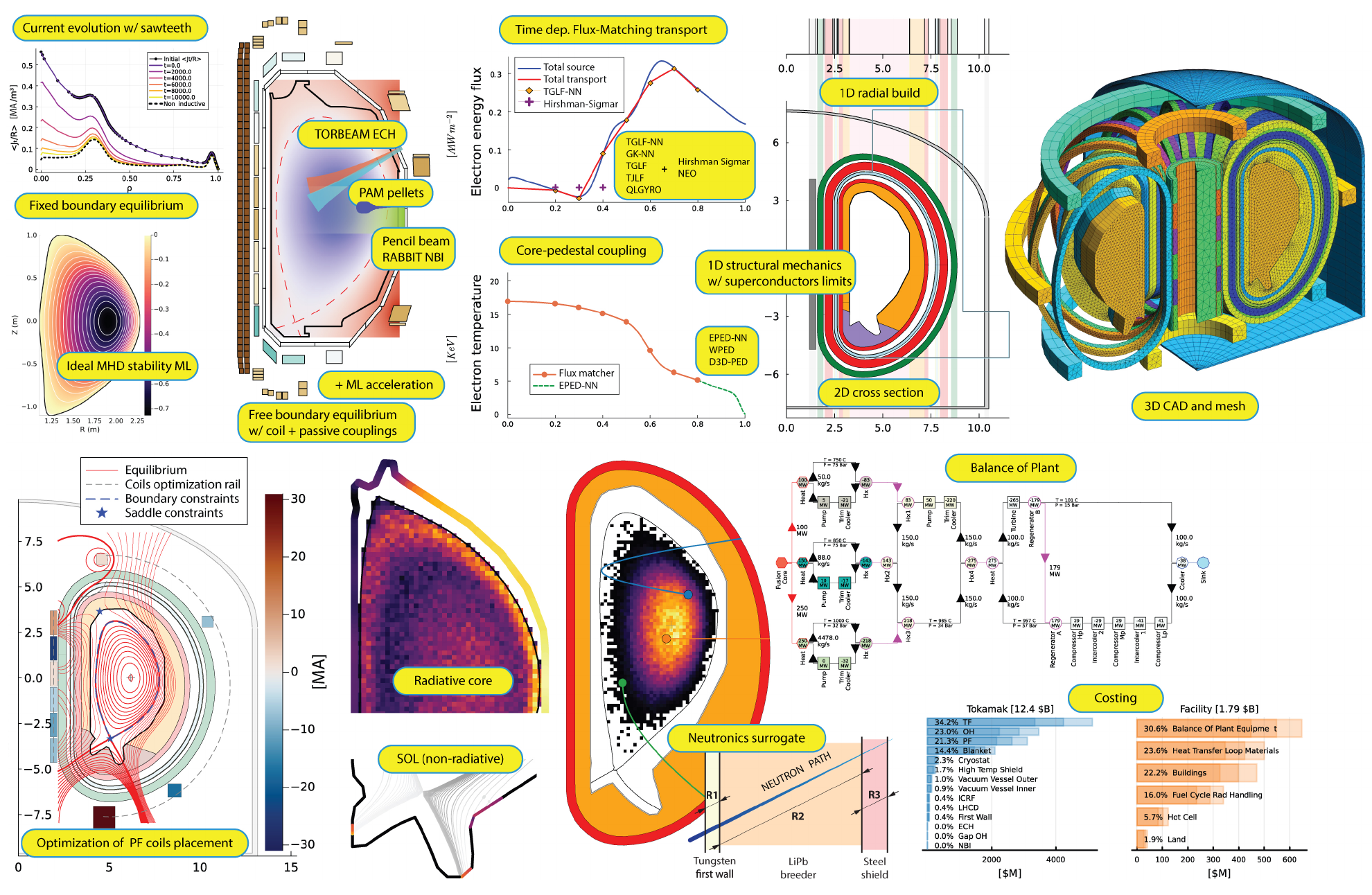}
    \caption{Representative outputs from a single FUSE whole facility evaluation \cite{meneghini2024fuse} demonstrating the breadth of integrated modeling. The framework seamlessly couples plasma physics with engineering design: \textbf{(Top left)} Current evolution and MHD equilibrium with optimized PF coil placement. \textbf{(Top center)} Core-pedestal transport coupling showing temperature and density profiles with heating and current drive sources. \textbf{(Top right)} Engineering design including 1D radial build, 2D cross-section, and 3D CAD model with mesh generation. \textbf{(Bottom)} Power exhaust modeling with radiative core and SOL physics, neutronics evaluation with surrogate models, balance of plant thermal cycle optimization, and direct capital cost breakdown. Each subsystem's outputs inform coupled analyses throughout the facility, ensuring self-consistent solutions that satisfy both physics requirements and engineering constraints. This comprehensive evaluation completes in 5-10 minutes, enabling the exploration of hundreds of thousands of design variants in optimization studies.}
    \label{fig:all_of_fuse}
\end{figure*}

FUSE couples plasma physics with engineering design through a unified computational environment, as illustrated in Figure~\ref{fig:all_of_fuse}. FUSE has the flexibility to incorporate configuration-specific physics models while maintaining consistent treatment of engineering subsystems. Both PT and NT designs are evaluated using identical costing models, engineering constraints, and optimization objectives. While FUSE captures comprehensive plasma physics and engineering constraints, vertical stability control is not yet included in the optimization framework. This is consistent with typical design practice where vertical control systems are refined after establishing reference plasma parameters \cite{humphreys_experimental_2009, nelson_implications_2024}. Post-optimization analysis using the TokaMaker code \cite{hansen_tokamaker_2024} confirmed that NT configurations exhibit higher vertical instability growth rates than PT, as expected from theory \cite{guizzo_assessment_2024} and DIII-D experiments \cite{nelson_vertical_2023}. While mitigation strategies exist, including passive stabilizing plates, we assume these can be implemented without fundamentally altering the conclusions.

\subsection{Plasma profile prediction: Adapting to configuration-specific physics}

Accurate prediction of plasma temperature and density profiles is essential for power plant design, as these profiles determine fusion power output, bootstrap current fraction, and exhaust power handling. However, the physics governing profile evolution differs between PT and NT, necessitating distinct modeling approaches.

For PT plasmas operating in H-mode, profile prediction follows well-established methodologies. Core transport modeling using theory-based models such as TGLF \cite{staebler2020geometry} (or neural network surrogates) provides temperature and density evolution from the magnetic axis to the pedestal region ($\rho \approx 0.8-0.9$). However, in the edge pedestal region, transport physics becomes inadequate as the profiles are instead constrained by magnetohydrodynamic (MHD) stability limits.

\begin{figure}[H]
    \centering
    \includegraphics[width=0.49\textwidth]{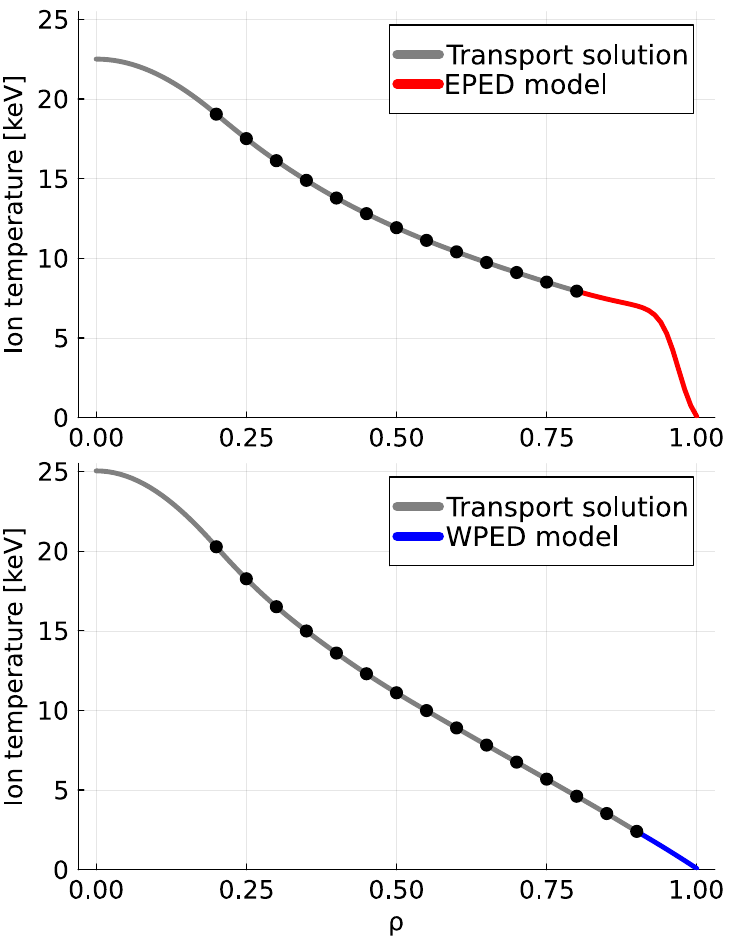}
    \caption{Schematic comparison of profile prediction methodologies for PT and NT configurations. \textbf{Top:} Positive triangularity H-mode with core transport predictions (black dots) extending to $\rho = 0.8$ and EPED model determining the steep pedestal structure (red). \textbf{Bottom:} Negative triangularity approach with transport predictions extending to $\rho = 0.9$ and WPED model providing weak edge gradients (blue). The absence of a strong pedestal in NT alters the modeling approach.}
    \label{fig:EPED_WPED}
\end{figure}

The EPED model \cite{snyder2011first, Snyder2009} captures this physics by predicting the pedestal height and width based on coupled peeling-ballooning and kinetic ballooning mode stability constraints. As shown schematically in Figure~\ref{fig:EPED_WPED} (top), this approach produces the characteristic steep pressure gradient that defines H-mode operation, provided the plasma power exceeds the L-H transition threshold \cite{Martin_2008}. This methodology has been validated across multiple devices \cite{slendebroek2023elevating}.

NT plasmas exhibit different edge physics. Extensive experimental campaigns~\cite{Camenen2005, pochelon_energy_1999, Austin2019, marinoni_brief_2021, nelson_robust_2023, thome_overview_2024, scotti_high_2024, nelson_first_2024, aucone_experiments_2024, paz-soldan_simultaneous_2024} demonstrate that strongly shaped NT configurations cannot access the peeling-ballooning stability space that enables H-mode pedestals. Instead, NT plasmas remain in a perpetual L-mode-like state with enhanced confinement. This offers the benefits of H-mode core performance without the liabilities of edge localized modes (ELMs). %

This absence of a distinct pedestal makes EPED inapplicable for NT \cite{nelson_characterization_2024}. Core transport models can extend closer to the edge ($\rho \approx 0.9$), but the outermost region still requires special treatment. To address this gap, we developed the Weak Pedestal (WPED) model based on systematic analysis of NT experimental data.

\subsection{The WPED model: A data-driven approach for NT edges}

The absence of a distinct pedestal in NT plasmas necessitates a different approach to edge modeling. Unlike PT configurations where steep gradients arise from MHD stability limits, NT edges exhibit smooth, continuous profiles that defy traditional pedestal models \cite{nelson_characterization_2024}. To capture this unique physics, we developed the WPED model based on systematic analysis of over 300 DIII-D NT discharges spanning a wide range of operational conditions \cite{thome_overview_2024}.

The model emerged from two robust empirical observations that persist across the entire NT database:

\begin{enumerate}
    \item \textbf{Constant energy fraction:} Despite variations in heating power, density, and impurity content, the ratio of thermal stored energy between edge ($\rho = 0.9-1.0$) and core ($\rho = 0-0.9$) regions remains remarkably constant \cite{nelson_characterization_2024}. This suggests a self-organizing process where the plasma naturally distributes energy to maintain a characteristic edge-to-core balance.
    
    \item \textbf{Smooth profile matching:} Temperature gradient scale lengths vary continuously across the core-edge boundary, contrasting sharply with the discontinuous jump characteristic of H-mode pedestals. This smooth transition indicates that transport processes, rather than stability limits, govern the edge structure.
\end{enumerate}

\begin{figure}[H]
    \centering
    \includegraphics[width=0.5\textwidth]{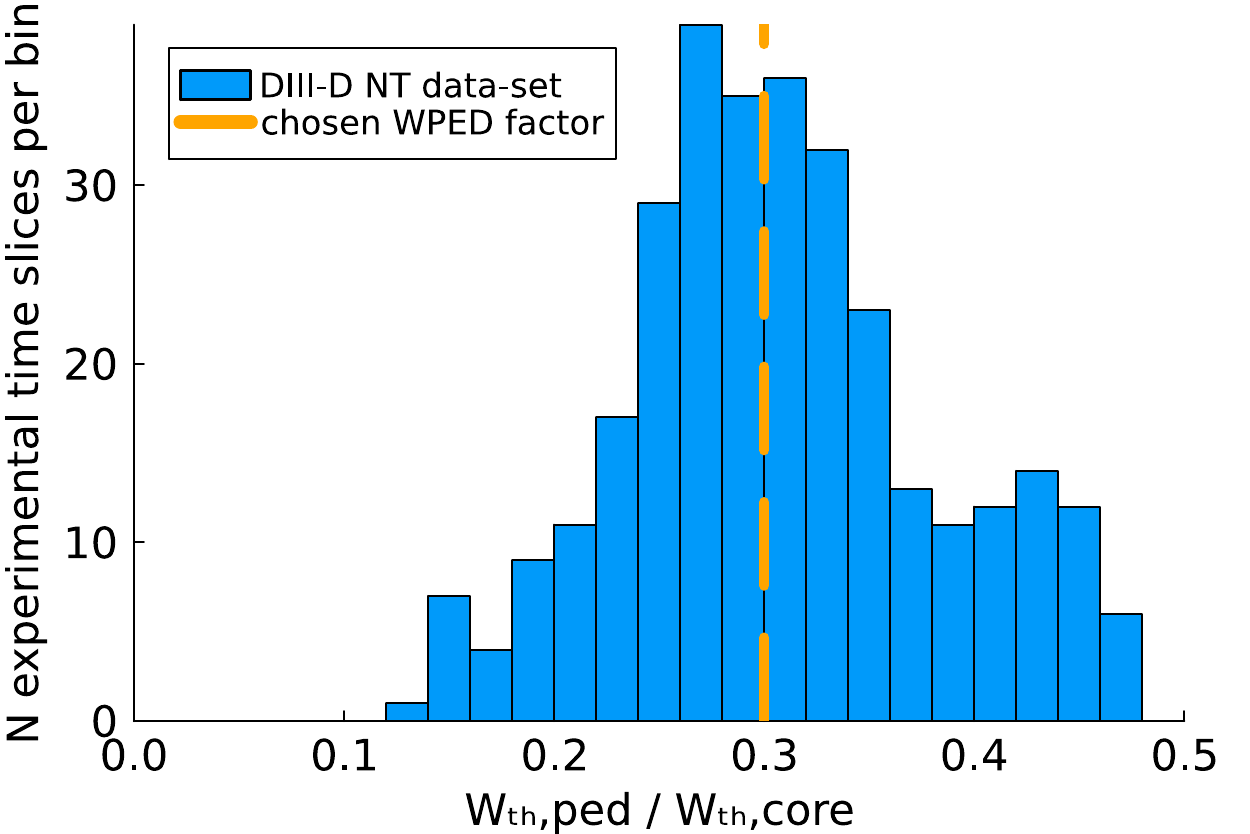}
    \caption{Distribution of thermal stored energy ratios between edge ($\rho = 0.9-1.0$) and core ($\rho = 0-0.9$) regions for 317 DIII-D NT time slices spanning 155 discharges. The peaked distribution around $c_{\text{WPED}} = 0.3$ demonstrates the empirical foundation for the WPED model's primary constraint. This consistency across varied plasma conditions—different powers, densities, and impurity levels—suggests a robust underlying physics mechanism rather than coincidental behavior.}
    \label{fig:wped_factor}
\end{figure}

Figure~\ref{fig:wped_factor} demonstrates the statistical foundation for the core-to-edge thermal stored energy constraint. The narrow distribution ($c_{\text{WPED}} = 0.30 \pm 0.05$) across hundreds of experimental time slices—spanning powers from 2-8 MW, densities from $2-6 \times 10^{19}$ m$^{-3}$, and various impurity scenarios—provides compelling evidence for this as a NT characteristic. Future work, especially comparison to ELM-free NT scenarios on other machines, is needed to support and refine this model. As discussed below, variations of this model are explored in this work as an initial substitute for a validated multi-facility NT dataset.

The WPED model translates these observations into a predictive framework through constrained optimization. The foundational principle is energy conservation: given a fixed edge-to-core energy ratio, the model determines edge profiles that simultaneously satisfy this constraint while maintaining smooth connection to core transport predictions \cite{nelson_characterization_2024, mcclenaghan_examining_2024}.

The energy ratio constraint is expressed mathematically as:

\begin{equation}
\frac{W_{\mathrm{th,ped}}}{W_{\mathrm{th,core}}}
  \;=\;
  \frac{\tfrac32\displaystyle\int_{\rho_{\mathrm{BC}}}^{1} P_{\mathrm{th}}(\rho)\,\mathrm dV}
       {\tfrac32\displaystyle\int_{0}^{\rho_{\mathrm{BC}}} P_{\mathrm{th}}(\rho)\,\mathrm dV}
  \;=\;
  c_{\mathrm{WPED}} ,
\label{equation:WPED}
\end{equation} where $W_{\mathrm{th,ped}}$ and $W_{\mathrm{th,core}}$ represent the thermal stored energies, $P_{\mathrm{th}}(\rho) = n_e T_e + \sum_i n_i T_i$ is the thermal pressure, and $\rho_{\mathrm{BC}} = 0.9$ marks the core-edge boundary. The choice of $\rho = 0.9$ as the transition point emerged from experimental analysis showing this location consistently marks where NT profiles begin deviating from core transport predictions.

To construct the edge profiles, WPED is based on exponential functions that naturally produce the smooth decay observed experimentally:

\begin{align}
T_e(\rho) &= T_{e,\mathrm{bound}} \exp\left[-\alpha_{T_e} \frac{(\rho - \rho_{\mathrm{BC}})}{(1 - \rho_{\mathrm{BC}})}\right], \quad \rho > \rho_{\mathrm{BC}} \\
T_i(\rho) &= T_{i,\mathrm{bound}} \exp\left[-\alpha_{T_i} \frac{(\rho - \rho_{\mathrm{BC}})}{(1 - \rho_{\mathrm{BC}})}\right], \quad \rho > \rho_{\mathrm{BC}}.
\end{align}
These functional forms offer several advantages: they guarantee positivity, provide smooth derivatives for gradient matching, and contain sufficient flexibility through the parameters $T_{e,\mathrm{bound}}$, $T_{i,\mathrm{bound}}$, $\alpha_{T_e}$, and $\alpha_{T_i}$ to match diverse experimental profiles.

Determining the four free parameters requires an optimization approach that balances multiple physics constraints. The WPED model uses a nested optimization scheme that mirrors the hierarchical nature of the constraints:

\begin{enumerate}
    \item \textbf{Outer optimization (Energy Conservation):} The boundary temperature $T_{e,\mathrm{bound}}$ is adjusted iteratively to satisfy the integral energy ratio constraint (Equation~\ref{equation:WPED}). This global constraint ensures the edge contains the appropriate fraction of total plasma energy.
    
    \item \textbf{Inner optimizations (Gradient Matching):} For each candidate $T_{e,\mathrm{bound}}$, the shape parameters $\alpha_{T_e}$ and $\alpha_{T_i}$ are optimized to minimize gradient scale length discontinuities at the core-edge boundary:
    
    \begin{align}
    \alpha_{T_e}^* &= \arg\min_{\alpha_{T_e}} \left|\frac{1}{L_{T_e}}\bigg|_{\rho_{\mathrm{BC}}^-} - \frac{1}{L_{T_e}}\bigg|_{\rho_{\mathrm{BC}}^+}\right|^2 \\
    \alpha_{T_i}^* &= \arg\min_{\alpha_{T_i}} \left|\frac{1}{L_{T_i}}\bigg|_{\rho_{\mathrm{BC}}^-} - \frac{1}{L_{T_i}}\bigg|_{\rho_{\mathrm{BC}}^+}\right|^2
    \end{align}
    
    where $L_T = -T/(\nabla T)$ is the temperature gradient scale length.
\end{enumerate}

The ion-electron temperature coupling is maintained through the constraint $T_{i,\mathrm{bound}} = T_{e,\mathrm{bound}} \cdot (T_i/T_e)|_{\mathrm{core}}$, preserving the core temperature ratio at the boundary. Separatrix conditions are enforced by setting $T_e(\rho=1) = T_{e,\mathrm{sep}}$ (typically 50-100 eV) and maintaining the edge temperature ratio.

The optimization uses the Brent's method \cite{gegenfurtner1992praxis} for its robustness in one-dimensional searches, resulting in sub-second computational times suitable for integration within the larger FUSE pedestal actor.

Sensitivity studies in \ref{sec:EXP_WPED} reveal the model's robustness: varying $c_{\mathrm{WPED}}$ by $\pm$50\% (encompassing the full experimental range) produces only $\pm$20\% variations in predicted edge temperatures. This insensitivity to the exact parameter value, combined with successful validation across diverse plasma conditions \cite{Neiser22} for the core transport, provides confidence in applying WPED to power plant-scale NT plasmas where direct experimental validation is impossible.

Having established the temperature profile methodology, we now describe the density profile treatment. Density profiles are determined through the optimization framework with boundary conditions set by the Greenwald density fraction at the pedestal region. For both configurations, the edge density is parameterized as $n_{e,ped} = f_{GW,ped} \times n_{GW}$, where nGW is the Greenwald density limit.
For PT configurations, the density boundary is set at the pedestal location determined by EPED, with the Greenwald fraction constrained to $0.3 < f_{GW,ped} < 1.0$, consistent with conventional H-mode operation limits observed across multiple tokamaks.
For NT configurations, the density boundary is set at $\rho = 0.9$ with an extended range of $0.3 < f_{GW,ped} < 1.3$. This broader range is justified by experimental observations showing NT plasmas can operate stably at higher density fractions without encountering the density limit disruptions typical of PT H-mode \cite{Sauter_2025}. The optimizer varies $f_{GW,ped}$ within these bounds to find optimal solutions while core density profiles evolve self-consistently through particle transport balance.

\subsection{Experimental validation: highly radiative Krypton seeding study}
\label{sec:EXP_WPED}
Model validation under power plant-relevant conditions is essential for confidence in optimization results. We tested WPED's predictive capability on DIII-D  discharge \#194306, a controlled krypton impurity seeding experiment that pushed NT plasmas to extreme radiative fractions approaching highly radiative power plant scenarios, conducted by Casali et al.\cite{casali2025achievement}.

The experiment's relevance extends beyond model validation: as revealed in our optimization studies as will be shown later in the results Section~\ref{sec:results}, NT power plant designs naturally operate with intense edge radiation due to reduced pedestal temperatures combined with significant impurity content. This radiative edge creates the large auxiliary power requirements identified for NT configurations while simultaneously providing inherent divertor protection.

Figure~\ref{fig:profile_validation} presents the comprehensive validation results through an animated comparison over the entire flattop (the plasma phase after the ramp-up where the total current is held steady). The discharge begins with moderate radiation ($f_{\text{rad,core}} \sim 0.2$) and progresses through increasing krypton injection to extreme radiative fractions ($f_{\text{rad,core}} > 0.7$) before approaching radiative collapse. Here we define the predicted $f_{rad,core} = \frac{P_{synchrotron}+P_{Kr,line}+P_{C,line}+P_{bremsstrahlung}}{P_{auxilary} + P_{ohmic}}$ where the core denotes the radiative power within the last closed flux surface.

\begin{figure}[H]
    \centering
    \animategraphics[loop,poster=1,width=0.5\textwidth]{1}{nt_comparison/frame_}{1}{13}    
    \caption{Validation of WPED model predictions against experimental measurements during controlled krypton seeding (DIII-D \#194306). The animation demonstrates model fidelity across the full range of radiative conditions relevant to NT power plant operation. \textbf{Top panels:} Electron and ion temperature profiles comparing experimental data (black symbols with error bars) to WPED predictions using different energy ratio parameters ($c_{\text{WPED}} = 0.2, 0.3, 0.4$). \textbf{Third panel:} Temporal evolution of edge temperatures showing prediction accuracy throughout the discharge. \textbf{Bottom panel:} Core radiation fraction predictions capturing the transition from moderate to extreme radiative conditions. The model's ability to track profile evolution during this challenging scenario validates its applicability to inherently radiative NT power plant designs.}
    \label{fig:profile_validation}
\end{figure}
Three key validation metrics demonstrate WPED's robustness. Temperature predictions remain within 10\% of experimental values for $\rho > 0.25$ across all radii and time points, even as radiation fractions triple during the discharge. While core temperature predictions ($\rho < 0.25$) show larger deviations of approximately 30\% between model and experiment, this discrepancy has limited impact on power plant design since the plasma volume and consequently fusion power production in this central region is negligible compared to the mid-radius region where fusion reactions peak. This edge model accuracy complements the validated core transport model discussed in Section~\ref{sec:integrated_modeling}, which achieves a mean relative error of 15\% for thermal stored energy across thousands of experimental discharges \cite{Neiser22}. The combination of accurate edge modeling (critical for power exhaust) and validated core transport (essential for fusion performance) provides confidence in the integrated plasma predictions. The model also shows remarkable insensitivity to parameter variations: changing $c_{\text{WPED}}$ from 0.2 to 0.4 produces only modest changes in predicted profiles, indicating the model is not overly dependent on the precise value of this empirical parameter, even though the core transport can be relatively stiff \cite{mcclenaghan_examining_2024}. Most importantly for power plant applications, WPED + TGLF accurately captures the evolution of the core radiation fraction $f_{\text{rad,core}}$, which is crucial for predicting auxiliary power requirements in power plant designs.

This validation provides confidence that WPED can reliably predict edge profiles for power plant design studies. For this validation, we utilized krypton impurity concentrations from the comprehensive AURORA transport analysis performed in Casali et al.~\cite{casali2025achievement}, interpolating between three reported time points to obtain continuous impurity profiles throughout the discharge. We took the on axis impurity fraction and used that fixed fraction for the entire profile. While this approach successfully captures the plasma response to established impurity distributions, we note that predictive modeling of impurity transport from active seeding remains an active area of research. Significant progress has been made in understanding impurity transport physics ~\cite{casson2014theoretical,odstrcil2017physics,sciortino2022investigation}. Future work coupling WPED with time-dependent impurity transport models will be essential for fully predictive simulations of radiative scenarios in NT power plants.

\section{Setting up design studies}
\label{sec:setup}

Having established the computational framework and edge physics models, we now turn to the specific optimization study designed to compare positive triangularity (PT) and negative triangularity (NT) configurations for fusion power plants. The comparison is structured as a CMOOP problem that systematically explores the design space to reveal fundamental trade-offs between economic viability and operational reliability.

\subsection{Optimization framework overview}

\begin{figure}
    \centering
    \includegraphics[width=0.5\textwidth]{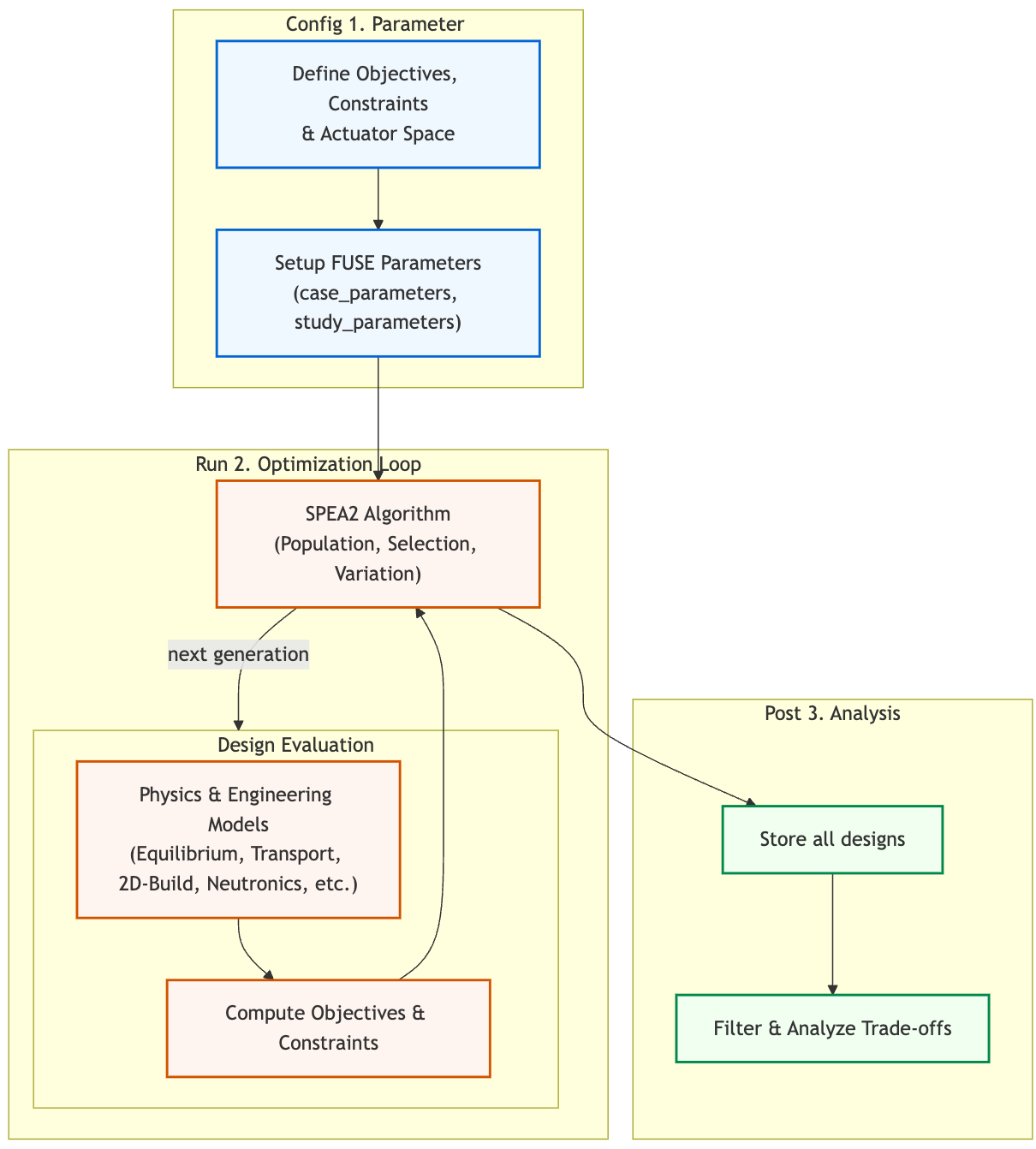}
    \caption{Schematic overview of the CMOOP study implemented in FUSE. The process begins with parameter configuration, where objectives (minimize capital cost, maximize $q_{95}$), constraints (engineering and physics limits), and actuator space (design variables) are defined. These are translated into FUSE parameters that feed into the optimization loop. The SPEA2 \cite{zitzler2001spea2} genetic algorithm evolves a population of designs through iterative evaluation, where each candidate undergoes complete physics and engineering assessment using FUSE's integrated models. The workflow stores all evaluated designs for post-processing analysis, enabling comprehensive exploration of design trade-offs and identification of Pareto-optimal solutions.}
    \label{fig:moop_flowchart}
\end{figure}

The optimization workflow, illustrated in Figure~\ref{fig:moop_flowchart}, leverages FUSE's integrated modeling capability to explore thousands of design variants while maintaining physical consistency across all subsystems. By physical consistency, we mean that each design satisfies fundamental conservation laws (energy, particle, and momentum balance), maintains continuous interface conditions between coupled domains (e.g., heat flux from core plasma through edge to divertor), and ensures that outputs from each physics module serve as self-consistent inputs to dependent subsystems. The process begins with parameter configuration, where we define competing objectives, engineering constraints, and the actuator space that encompasses both continuous variables (e.g., major radius, magnetic field) and discrete choices (e.g., superconductor technology, balance of plant cycle types). These parameters are translated into FUSE's comprehensive data structure (IMAS \cite{imbeaux2015design} + extras built on top of IMAS), which propagates through all coupled physics and engineering modules during each design evaluation.

The driving force of the CMOOP study is the Strength Pareto Evolutionary Algorithm 2(SPEA2) \cite{zitzler2001spea2} genetic algorithm, implemented in the Julia package Metaheuristics.jl \cite{metaheuristics2022}. We empirically tested several multi-objective optimization algorithms, including NSGA-II \cite{996017}, SMS-EMOA, and CCMO, on our fusion design problem. This revealed significant performance differences. NSGA-II, despite its widespread adoption, converged to limited regions of the Pareto front rather than exploring the full trade-off space. The Pareto front denotes the set of designs where improving one objective degrades the other. SMS-EMOA exhibited convergence difficulties for our constrained problem, while CCMO showed no improvement over SPEA2. In contrast, SPEA2 \cite{zitzler2001spea2} demonstrated both robust convergence and comprehensive Pareto front coverage, efficiently exploring the multi-objective design space. Unlike traditional optimization methods that require objective weighting, SPEA2 maintains a diverse population of solutions that evolve toward the Pareto front. This approach is particularly valuable for fusion power plant design, where the trade-offs between cost and performance are not known \textit{a priori} and may be highly nonlinear.
\subsection{Multi-objective formulation}

The optimization seeks to balance two competing objectives that capture the fundamental tension in fusion power plant design:

\begin{enumerate}
    \item \textbf{Minimize capital cost}: This drives the design toward compact, economically attractive solutions with smaller magnets, reduced structural materials, and lower overall construction costs. We utilize the ARIES costing model \cite{waganer2013aries} in this study, and emphasize that several assumptions (eg. the cost of various coil technologies) are fixed with potentially large uncertainties. As such, only relative values and not absolute dollar values should be extracted from this work. Regardless, the minimization of capital cost serves as a useful objective to help guide the optimization process. 
    
    \item \textbf{Maximize safety factor $q_{95}$}: This reduces operational risk by increasing the margin against disruptions. Evidence from PT DIII-D experiments demonstrates that disruption rates decrease significantly with increasing $q_{95}$, with per-discharge disruptivity dropping from over 80\% at $q_{95} \sim 2$ to below 10\% at $q_{95} > 5$ in a database of approximately 6000 PT discharges~\cite{garofalo2014fusion}. This strong dependence on $q_{95}$ makes maximizing the safety factor a critical design consideration for reliable power plant operation.
\end{enumerate}

We deliberately chose direct capital cost rather than levelized cost of electricity to focus on near-term engineering and construction challenges without introducing uncertainties about component lifetimes, capacity factors, and operational scenarios for which limited data exists. The emphasis on $q_{95}$ reflects the critical importance of disruption avoidance for power plant viability, as disruptions can cause severe damage to plasma-facing components and limit plant availability. Because robust data detailing the disruptivity of NT plasmas is not yet available we treat NT and PT plasmas the same while applying this objective. However, NT discharges on DIII-D were able to routinely achieve high confinement at low $q_{95}<3$, which could lead to a widening of the fusion power plant design space \cite{paz-soldan_simultaneous_2024}.

The optimization is subject to constraints that ensure consistency with established plant requirements while retaining space for engineering risk. For net electric power output, we require $P_{\mathrm{el,net}} = 250 \pm 50$ MW$_e$, placing designs in the upper half of the 40–200 MW$_e$ window recommended by the NASEM plant study for market-relevant scale~\cite{nasem2021}. Fuel self-sufficiency demands a tritium breeding ratio exceeding 1.1, providing a 20\% margin over the NASEM minimum of 0.9 to accommodate modeling and inventory uncertainties~\cite{nasem2021}. In addition, we require that all designs be able to sustain at least one-hour flattop, matching the historical FESAC ``500 MW$_{\text{fus}}$ for 1 h" criterion~\cite{fesac2003}. Additional optimization of the pulse length and availability will be the focus of future work.

Engineering constraints ensure component integrity and operational viability. The superconducting magnet current density and von Mises stress must remain below material limits to maintain conservative structural and thermal margins. We adopt the power exhaust benchmark of $P_{\text{sol}}/R < 15$ MW/m derived from Kallenbach et al. \cite{kallenbach2013impurity}, which synthesizes experimental results from current devices and projects the heat-handling requirements for ITER and DEMO. This constraint represents an estimate for sustainable tungsten divertor operation, beyond which material erosion and thermal damage would compromise steady-state performance \cite{kallenbach2013impurity}. Additionally, PT configurations must satisfy $P_{\mathrm{sol}}/P_{LH} > 1.1$  to ensure H-mode access, providing a 10\% buffer above the L-H threshold. We implemented the threshold scaling as described in \cite{schmitz2022reducing}, including the metal wall and isotope effects from \cite{birkenmeier2022power}. This constraint does not apply to NT configurations due to their inherently ELM-free operation.

Global MHD stability considerations are implicitly satisfied through our operational approach rather than explicit constraints. Our designs utilize fully relaxed current profiles evolved primarily by bootstrap current rather than extensive auxiliary current drive shaping. Combined with moderate fueling scenarios that avoid large pellet-induced pressure perturbations, this approach naturally maintains operation below the no-wall $\beta_N$ limit. Indeed, all feasible designs in our optimization database are below $\beta_N < 1.7$ without requiring this as an explicit constraint, demonstrating that conservative profile evolution strategies inherently provide MHD stability margins.

\subsection{Design space and actuators}

Table~\ref{tab:objectives_constraints_actuators} summarizes the complete optimization problem, including the actuator space that defines the range of design parameters explored. The optimization considers both tokamak geometry (major radius 4.5-10 m, magnetic field 8-15 T, plasma current 10-22 MA) and operational parameters (density, impurity content, auxiliary heating). These ranges were chosen based on earlier studies to be sufficiently broad that no feasible designs reach the boundaries. This approach accelerates convergence without constraining the optimization outcomes. The triangularity parameter critically differentiates the two configurations: $\delta = 0.3$ to 0.7 for PT versus $\delta = -0.3$ to -0.7 for NT, with both configurations maintaining elongation at $\kappa = 1.8$. Throughout this study we fix the aspect ratio at $R_{0}/a = 3.24$, slightly above which most contemporary tokamaks (e.g., DIII-D, JET, ASDEX Upgrade) have been designed and for which the underlying physics and engineering models are most validated. This aspect ratio also provides practical engineering advantages, allowing sufficient radial space for both the inboard blanket and toroidal field (TF) coils without excessive compression. Future optimization studies could explore the full spectrum from spherical tokamaks (aspect ratio $\sim$1.5-2.0) to large aspect ratio devices ($>$4.0), though spherical tokamak designs may require removal of the inboard blanket to accommodate the constrained radial build, fundamentally altering the tritium breeding and neutron shielding strategy.

\begin{table*}[!ht]
\centering
\begin{tabular}{ccc}
    \begin{minipage}{0.26\textwidth}
        \centering
        \begin{tabular}{|c|}
        \hline
        \textbf{Objectives} \\ \hline\hline
        Minimize capital cost \\ \hline
        Maximize $q_{95}$ \\ \hline
        \end{tabular}
    \end{minipage}
    &
    \begin{minipage}{0.44\textwidth}
        \centering
        \begin{tabular}{|c|}
        \hline
        \textbf{Constraints} \\ \hline\hline
        $P_{\text{electric}} = 250 \pm 50$ [MWe] \\ \hline
        flattop time $= 1$ [hr] \\ \hline
        TBR $= 1.1 $ \\ \hline
        \textcolor{red}{$P_{\text{sol}} / P_{LH} > 1.1$} \\ \hline
        $P_{\text{sol}} / R < 15$ [MW/m] \\ \hline
        $j_{\text{crit,TF}}$, $j_{\text{crit,OH}}$, $j_{\text{crit,PF}} < j_{\text{crit,material}}$ \\ \hline
        $\sigma_{\text{crit,TF}}$, $\sigma_{\text{crit,OH}}$, $\sigma_{\text{crit,PF}} < \sigma_{\text{material}}$ \\ \hline
        $\epsilon_{transport} < 1.0$ \\ \hline
        $q_{95} > 3.0$ \\ \hline
        \end{tabular}
    \end{minipage}
    &
    \begin{minipage}{0.30\textwidth}
        \centering
        \begin{tabular}{|c|}
        \hline
        \textbf{Actuators} \\ \hline\hline
        $4.5 < R_0 < 10.0$ [m] \\ \hline
        $8.0 < B_0 < 15.0$ [T] \\ \hline
        $10.0 < I_p < 22.0$ [MA] \\ \hline
        \textcolor{blue}{$-0.7 < \delta < -0.3$} \\ \hline
        \textcolor{red}{$0.0 < \delta < 0.7$} \\ \hline
        \textcolor{red}{$1.5 < Z_{\text{eff, ped}} < 2.5$} \\ \hline
        $0.2 < f_{GW, \text{ped}} < 1.3$ \\ \hline
        $0 < P_{EC} < 50$ [MW] \\ \hline
        $0 < \rho_{EC} < 0.9$ \\ \hline
        $0 < P_{IC} < 50$ [MW] \\ \hline
        TF shapes \\ \hline
        \end{tabular}
    \end{minipage}
\end{tabular}
\caption{Objectives, constraints, and actuators for the fusion power plant multi-objective constrained optimization. Red items apply only to positive triangularity (PT) configurations, while blue items apply only to negative triangularity (NT). The L-H power threshold constraint (red) is not applicable to NT due to its inherently ELM-free operation, while edge impurity content $Z_{\text{eff,ped}}$
is bounded at a minimum of 1.5 for both configurations but allowed to increase further for PT to enable radiative solutions.}
\label{tab:objectives_constraints_actuators}
\end{table*}

Beyond geometric parameters, the optimization explores auxiliary heating scenarios with up to 100 MW combined electron cyclotron (EC) and ion cyclotron (IC) power, with optimizable deposition locations. The TF coil design space includes three cross-sectional shapes (racetrack, double ellipse, offset) and both low-temperature (Nb$_3$Sn) and high-temperature (ReBCO) superconductor technologies, allowing assessment of how advanced magnet technology impacts the optimal design point.

\subsection{Integrated Modeling Workflow}
\label{sec:integrated_modeling}
The optimization studies employ FUSE's hierarchical actor system to ensure self-consistent solutions across all physics and engineering domains. The workflow proceeds through two primary stages: plasma equilibration and facility-wide integration. Detailed descriptions of the individual actors and their implementation can be found in the FUSE reference~\cite{meneghini2024fuse}.

The \texttt{ActorStationaryPlasma} handles iterative convergence between plasma equilibrium, current drive, transport, and heating systems. Starting from initial profiles, the workflow begins with \texttt{ActorHCD} calculating electron cyclotron (EC) and ion cyclotron (IC) heating deposition profiles, where EC power deposits between $\rho = 0.2$--$0.8$ (optimizer-controlled) while IC deposits at $\rho = 0.1$. For this study, we employ a simplified heating absorption model similar to that described in \cite{slendebroek2023elevating}. This approach is justified because our designs operate as strongly burning plasmas at flattop, with fusion alpha heating dominating auxiliary heating by a factor of 10:1. The \texttt{ActorCurrent} module with \texttt{ActorQED} \cite{Lyons24} then updates the plasma current density $J_{tor}(\rho)$ accounting for Ohmic, bootstrap, and auxiliary current drive contributions, allowing the current profile to naturally evolve rather than being actively tailored. While this simplified treatment suffices for steady-state analysis, we note that future work addressing plasma startup scenarios will require time-dependent modeling with more detailed auxiliary heating and current drive calculations, capabilities currently under rapid development within FUSE.

Edge boundary conditions are established through \texttt{ActorPedestal}, which applies configuration-specific models---EPED for PT or WPED for NT---to set edge temperatures and densities at $\rho = 0.8$--$1.0$. The core plasma profiles evolve through \texttt{ActorCoreTransport} with \texttt{ActorFluxMatcher}, which advances $T_e(\rho)$, $T_i(\rho)$, and $n_e(\rho)$ using  TGLF neural network surrogates \cite{Neiser22}. Both configurations use a neural net based on the SAT0 saturation rule including EM fluctuations model trained on multi-machine databases including electromagnetic effects. This model has been validated on NT, PT, and L-mode plasmas with a mean relative error (MRE) of 15\% for the thermal stored energy across 3000 H-mode, 1500 L-mode and 250 NT DIII-D experimental time slices \cite{Neiser22}. Spot checks comparing the neural network to full TGLF calculations confirm agreement within model uncertainties. While the newer SAT3 saturation rule with GKNN \cite{Neiser23} correction terms demonstrates improved accuracy (MRE of 11\% on the same dataset), it remains under development for power plant-scale applications as the GKNN addition is very sensitive to the training's data range. The transport model extends to $\rho = 0.8$ for PT and $\rho = 0.9$ for NT configurations.

The last step of the \texttt{ActorStationaryPlasma} is \texttt{ActorEquilibrium} with \texttt{ActorTEQUILA} \cite{Lyons24}, a fixed boundary equilibrium code that solves the Grad-Shafranov equation using the updated pressure and current profiles. This cycle repeats until convergence, defined as $< 5\%$ relative change in current and pressure profiles between iterations, typically requiring 3--5 iterations.

From the viewpoint of transport analysis both configurations assume no reduction in turbulent fluxes from rotation shearing $(d\omega_0/d\rho=0)$. The impurity content is Kr-He with a constant helium fraction of $f_{He}=0.01$. The Krypton fraction $f_{Kr}$ is adjusted to match the specified $Z_{\text{eff}}$ profile while maintaining quasi-neutrality with 50:50 D-T fuel mixture.

Once plasma convergence is achieved, \texttt{ActorWholeFacility} handles the complete plant design through a series of interconnected analyses. The magnet system design begins with \texttt{ActorHFSsizing} and \texttt{ActorLFSsizing} optimizing TF coil geometry while considering stress limits, superconductor critical currents, and ripple constraints. Subsequently, \texttt{ActorPFdesign} positions the poloidal field coils to maintain equilibrium shape control throughout the discharge while minimizing stored magnetic energy.

Nuclear systems analysis proceeds through \texttt{ActorNeutronics} calculating neutron wall loading, followed by \texttt{ActorBlanket} optimizing blanket thickness to achieve the required tritium breeding ratio exceeding 1.1. Power systems evaluation encompasses \texttt{ActorDivertors} for heat exhaust handling and \texttt{ActorBalanceOfPlant} for thermal conversion cycle optimization. Finally, \texttt{ActorCosting} computes direct capital costs using the ARIES \cite{waganer2013aries} model with inflation adjustment to present-day dollars. These estimates represent Nth-of-a-kind fusion power plants rather than first-of-a-kind demonstrations. Because of uncertainty in the cost estimation, \texttt{ActorCosting} is best used for comparing relative power plant costs instead of the absolute dollar amount.

This hierarchical approach ensures that modifications in any subsystem propagate consistently throughout the design. For instance, changes to the magnetic field strength affect not only confinement but also mechanical stresses, blanket thickness requirements, and ultimately plant economics.

\subsection{Optimization execution and convergence}

Each optimization study evolves a population of 320 designs over 50 generations using the SPEA2 \cite{zitzler2001spea2} algorithm, totaling approximately 16,000 full facility evaluations per study. Individual evaluations require 5-10 minutes per design point to execute FUSE's complete \texttt{ActorWholeFacility}. Distributed computing across 160 parallel workers on a moderate HPC cluster with 320 cores enables study completion in 10-20 hours.

\begin{figure}[H]
    \centering
    \includegraphics[width=0.49\textwidth]{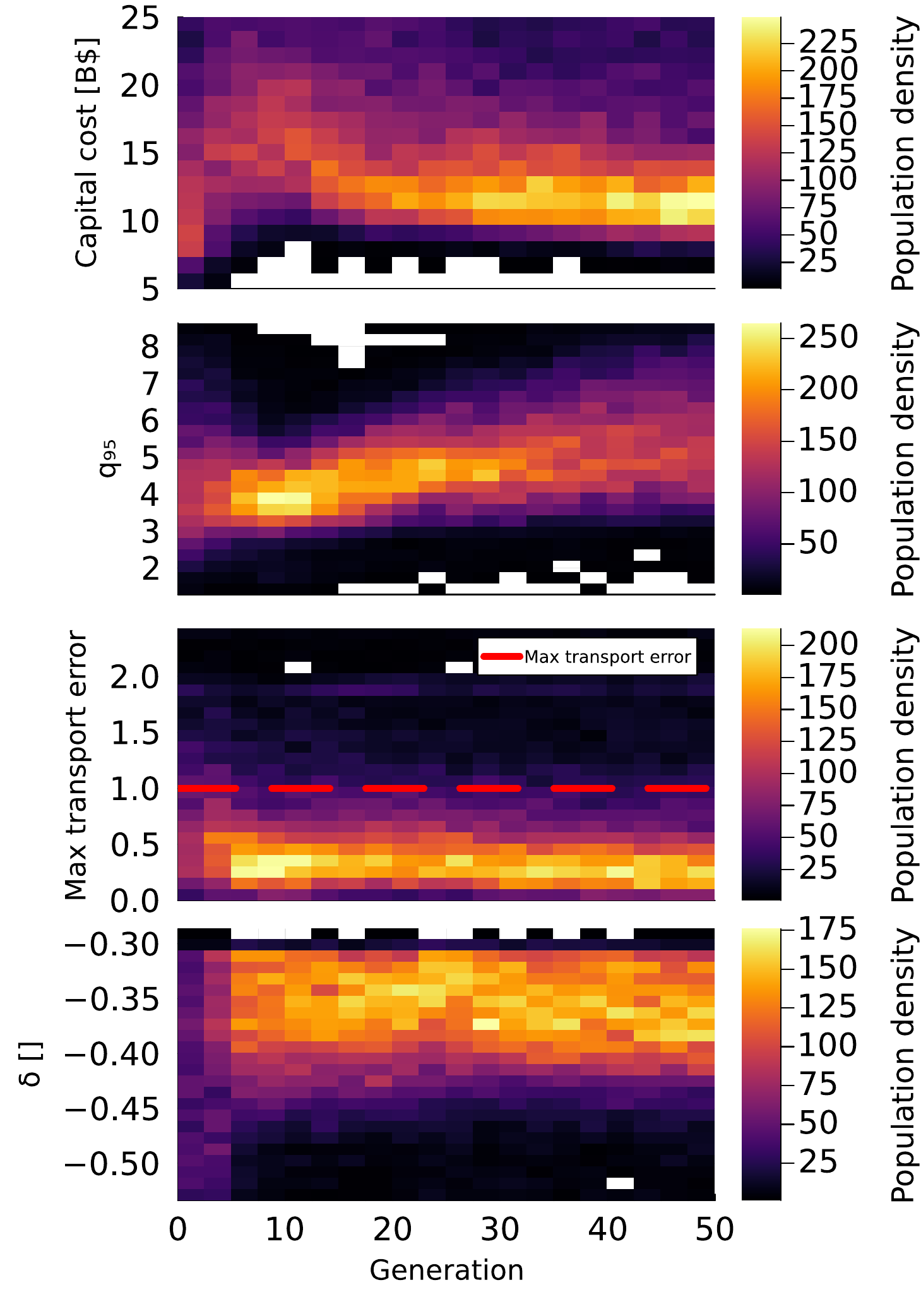}
    \caption{Evolution of the genetic algorithm population across 50 generations showing the full explored design space (not filtered by constraints). \textbf{First and second panel}: Objective functions showing capital cost and $q_{95}$ distributions. Early generations explore broadly while later generations concentrate on Pareto-optimal regions. \textbf{Third panel}: Transport convergence error constraint, with the red line indicating the maximum allowable error ($<1.0$) for valid solutions. \textbf{Bottom panel}: Triangularity actuator evolution for NT optimization, demonstrating the optimizer's preference for weakly negative values (-0.3 to -0.4). The population density (color scale) reveals the algorithm's exploration strategy, with white regions indicating parameter combinations that are either unexplored or lead to constraint violations.}
    \label{fig:moop_traces}
\end{figure}

Figure~\ref{fig:moop_traces} reveals how the genetic algorithm explores and refines the design space over successive generations. The top panels show the evolution of the two objectives: capital cost and $q_{95}$. Early generations (blue) explore broadly, testing extreme parameter combinations to map the feasible space. As evolution progresses (yellow), the population concentrates on promising regions that balance both objectives. The transport convergence panel demonstrates that the algorithm successfully identifies designs satisfying the convergence criteria ($\epsilon_{\text{transport}} < 1.0$), with failed designs naturally eliminated through selection pressure.

Most notably, the triangularity evolution (bottom panel) shows the optimizer consistently drives NT designs toward weakly negative values near $-0.4<\delta<-0.3$, suggesting this provides the best balance of confinement performance and engineering feasibility. This convergent behavior validates both the optimization methodology and the physics models, as the algorithm independently discovers design features consistent with experimental observations.

The emergence of well-defined Pareto fronts is visualized in Figure~\ref{fig:pareto_animation}, which animates the population evolution for NT configurations. Initially scattered designs gradually organize into a clear trade-off curve, revealing that achieving higher $q_{95}$ values (reduced disruption risk) requires progressively larger and more expensive machines. The smooth final Pareto front indicates good convergence and suggests our actuator space captures the key design variables. A complete statistical breakdown of all designs that met these constraints is provided in \ref{apendix:Design_space_statistics}, Table \ref{tab:results_all_gen_combined}, for ease of comparison with traditional systems studies.

\begin{figure}[H]
    \centering
    \animategraphics[loop,poster=44,width=0.5\textwidth]{8}{MOOP_results/par_anim_WPED_s/frame_}{1}{45}
    \caption{Evolution of the Pareto front for negative triangularity (NT) configurations across 50 generations of the genetic algorithm optimization. The animation demonstrates convergence from an initially scattered population toward the Pareto-optimal trade-off between capital cost (y-axis) and safety factor $q_{95}$ (x-axis). Color indicates generation number, progressing from early exploration (purple) to final convergence (yellow). The emerging Pareto front reveals the fundamental cost-risk trade-off: higher $q_{95}$ values that reduce disruption risk require larger, more expensive machines. The smooth convergence and well-defined front validate both the optimization methodology and the underlying physics models. Click to view the animation (requires PDF readers with animation support such as Adobe Acrobat).}
    \label{fig:pareto_animation}    
\end{figure}
Additional optimization studies were conducted to explore the sensitivity of these conclusions to key assumptions, including power exhaust limits ($P_{\text{sol}}/R = 15$ or 30 MW/m), pedestal performance ($c_{\text{EPED}} = 0.6, 0.8, 1.0$ for PT), L-H transition uncertainty, and superconductor technology. This comprehensive exploration, totaling over 200,000 integrated design evaluations, provides robust statistics for comparing PT and NT configurations across a range of physics and technology assumptions.

\section{Fusion power plant optimization}
\label{sec:results}

\subsection{Pareto Front Analysis and Configuration Comparison}

\begin{figure}[H]
    \centering
    \includegraphics[width=0.49\textwidth]{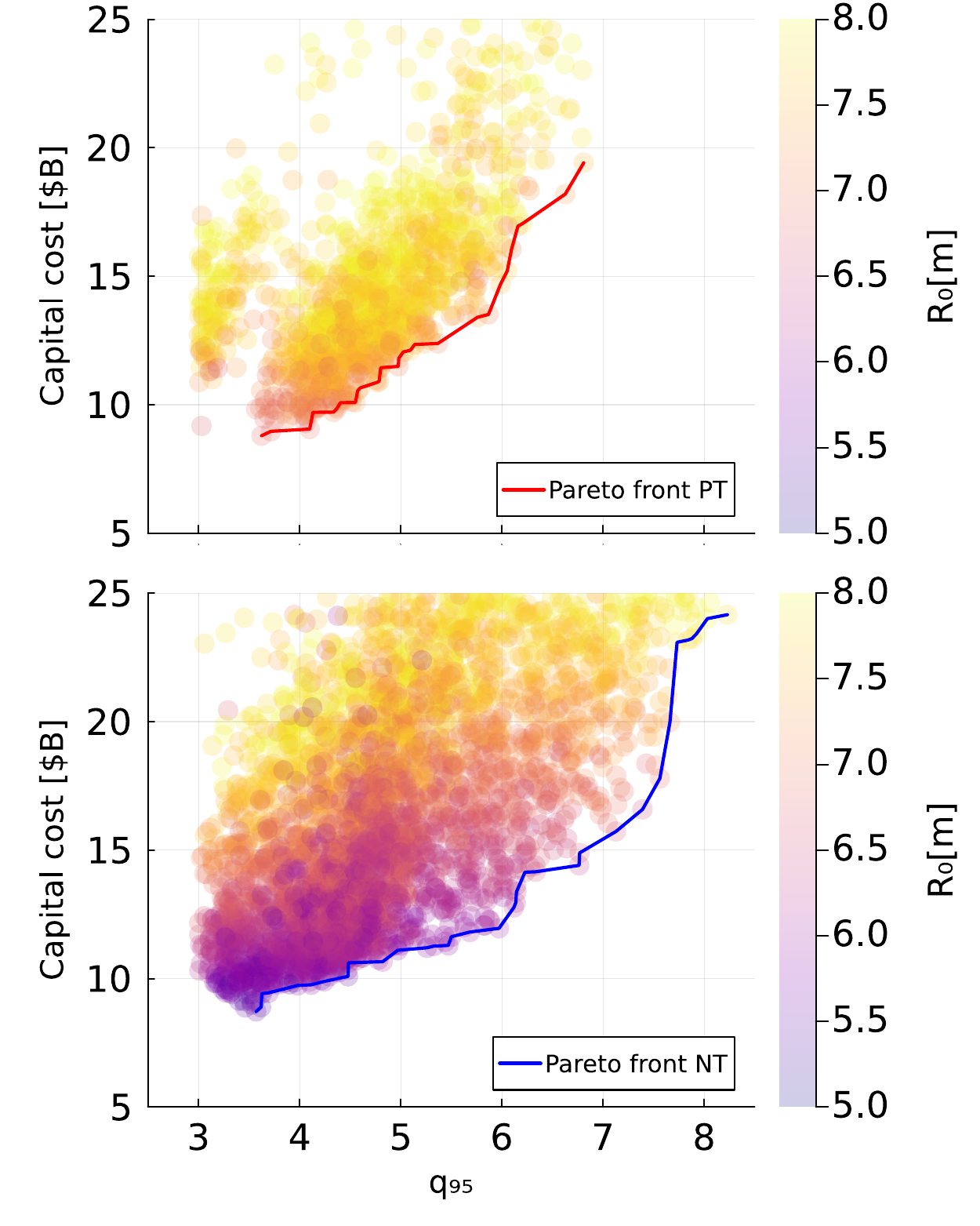}
    \caption{Pareto-optimal fusion power plant designs revealing distinct optimization strategies for PT and NT configurations. Color indicates major radius $R_0$. \textbf{Top}: PT configurations are constrained to larger machines ($R_0 > 6.5$ m, yellow) due to power exhaust limits, achieving cost optimization through reduced magnetic field strength ($B_0 \sim 8$ T). \textbf{Bottom}: NT configurations access compact, high-field designs ($R_0 \sim 5.5$ m, purple) that achieve cost parity with PT through $B_0 > 12$ T operation. The extended NT Pareto front at high $q_{95}$ demonstrates greater operational flexibility, while both configurations show similar cost scaling with safety factor despite their contrasting design philosophies.}
    \label{fig:PT_pareto}
\end{figure}

The fundamental trade-off between capital cost and operational reliability emerges clearly in the Pareto fronts shown in Figure~\ref{fig:PT_pareto}. The Pareto front denotes the set of designs where improving one objective degrades the other; for two objectives this maps out a 2D line. Despite their different physics constraints and operational characteristics, both PT and NT configurations achieve remarkably similar cost-performance trade-offs, but through different design strategies. While the Pareto front figures focus on optimal fronts, \ref{apendix:Design_space_statistics} (Table \ref{tab:results_all_gen_combined}) reports $\mu \pm \sigma$ values for every key variable across the entire feasible population, mirroring the summary style used in classical systems-code studies.

The most striking feature is the systematic difference in machine size between configurations. PT designs cluster at major radii above 6.5 meters (appearing yellow in the figure), forced to larger major radii by the need to distribute power exhaust below the $P_{\text{sol}}/R$ limit while maintaining $P_{sol} > P_{LH}$. To minimize costs at these larger sizes, the optimizer reduces the magnetic field to approximately 8 T, reducing the cost of the most expensive item for the power plant.

In contrast, NT configurations populate the compact design space with major radii as low as 5.5 meters (purple regions), enabled by their ability to radiate power in the edge region and not being constraint by the L-H power threshold. These compact NT machines require magnetic fields exceeding 11 T on axis to achieve adequate confinement without a strong pedestal, explaining their natural synergy with HTS technology discussed in Section~\ref{sec:HTS_LTS}. Despite requiring approximately 30\% smaller major radius and 50\% higher magnetic field, NT achieves similar capital costs to PT through this high-field approach.

The extended range of the NT Pareto front, particularly at high $q_{95}$ values exceeding 8, reflects its greater operational flexibility. While PT configurations become increasingly constrained by the competing requirements of H-mode access and power exhaust management at high safety factors, NT can continue optimizing toward lower risk operation. This could enable a risk-averse power plant design which increases power plant availability at a minimal capital cost increase.

These results challenge the conventional wisdom that NT inherently requires larger, more expensive machines than PT. Instead, our optimization reveals that NT naturally evolves toward a more compact, high-field design philosophy that achieves economic equivalence with conventional PT approaches.

\subsection{Core-Edge Power Coupling and Radiative Scenarios}
\begin{figure*}[ht]
    \centering
    \includegraphics[width=\textwidth]{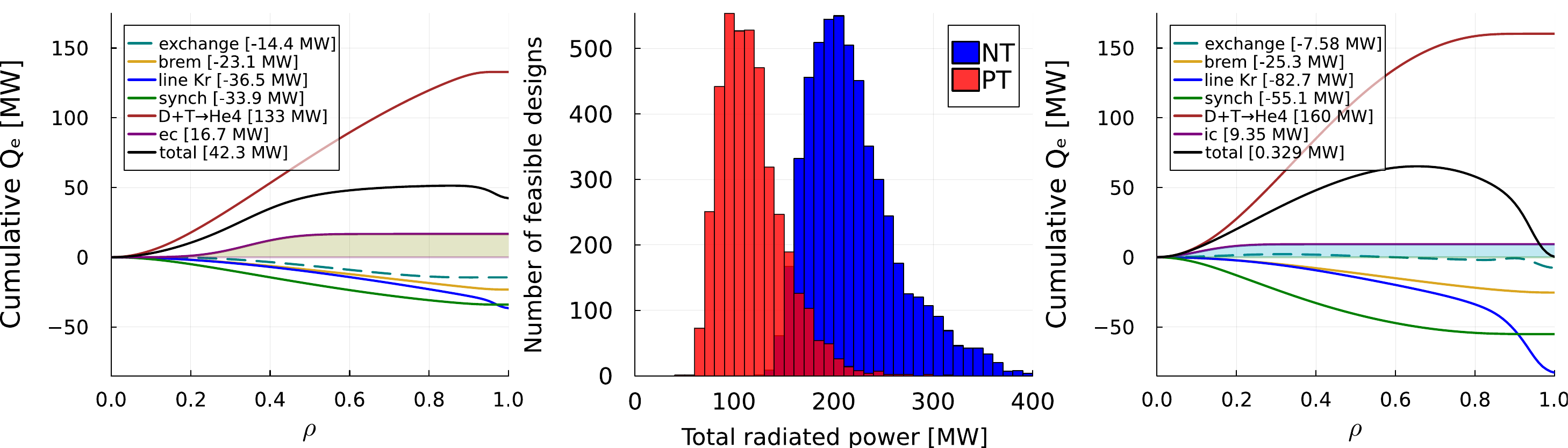}
    \caption{Power balance and radiation characteristics for representative PT and NT designs from the Pareto front. \textbf{Left}: PT configuration showing cumulative electron energy sources with total integrated power of 42.3 MW. Fusion heating dominates in the core ($\rho = 0.2-0.8$) with modest auxiliary heating (shaded components) and minimal edge radiation losses. \textbf{Center}: Distribution of total radiated power across the design database, demonstrating NT configurations naturally operate with approximately double the radiation fraction of PT designs. \textbf{Right}: NT configuration with dramatically different power balance, showing only 0.33 MW total integrated power due to intense edge radiation losses that nearly cancel the core heating. All profiles shown at flattop; startup auxiliary power requirements are left for future work.}
    \label{fig:radiation_profiles}
\end{figure*}

A fundamental difference between PT and NT configurations emerges in their power balance and core-edge coupling, as illustrated in Figure~\ref{fig:radiation_profiles}. The cumulative electron energy source profiles reveal strikingly different radiation patterns that drive the exhaust solutions for each configuration.

For PT configurations (left panel), the cumulative power profile shows classical behavior: fusion heating (D+T→He4) provides the dominant contribution, peaking in the near core where temperatures and the volume are the highest. Radiation losses from krypton impurity seeding and other mechanisms remain modest, allowing the configuration to maintain a positive integrated power of 42.3 MW that flows to the edge for exhaust through the divertor.

NT configurations (right panel) exhibit fundamentally different behavior. Despite similar fusion power generation in the near core, the combination of reduced edge temperatures and enhanced impurity radiation from krypton at lower edge temperatures creates severe radiative losses approaching 80 MW at the edge ($\rho > 0.9$). This edge radiation sink effectively cancels most of the core power, resulting in a near-zero total integrated power of only 0.33 MW.

The center histogram quantifies this effect across the full design database: NT configurations typically radiate 200-250 MW compared to 80-120 MW for PT designs. This factor-of-two difference in radiated power fraction explains the need for a slightly increased fusion power source and validates the choice of highly radiative DIII-D discharges for WPED model validation (Section~\ref{sec:EXP_WPED}). The enhanced radiation also provides inherent exhaust handling advantages, distributing the power load over the entire first wall rather than concentrating it in the divertor.

This radiative core-edge coupling represents both a challenge and an opportunity for NT power plant design. Future work should address start-up auxiliary heating requirements and power ramp strategies. Moreover, the intrinsic radiative dissipation may eliminate the need for advanced divertor solutions required for PT configurations, potentially offsetting the increased heating system costs through simplified heat exhaust management.

We note that steady-state advanced tokamak scenarios, including those with negative triangularity configurations optimized for high bootstrap fraction operation with reversed magnetic shear \cite{ZHENG2024100051}, represent an important complementary design approach. Such AT scenarios, while offering potential advantages in normalized beta limits, involve different operational considerations and stability constraints beyond the scope of this power plant optimization study.

\subsection{Power Exhaust Constraints and Operational Windows}

\begin{figure}[H]
    \centering
    \includegraphics[width=0.49\textwidth]{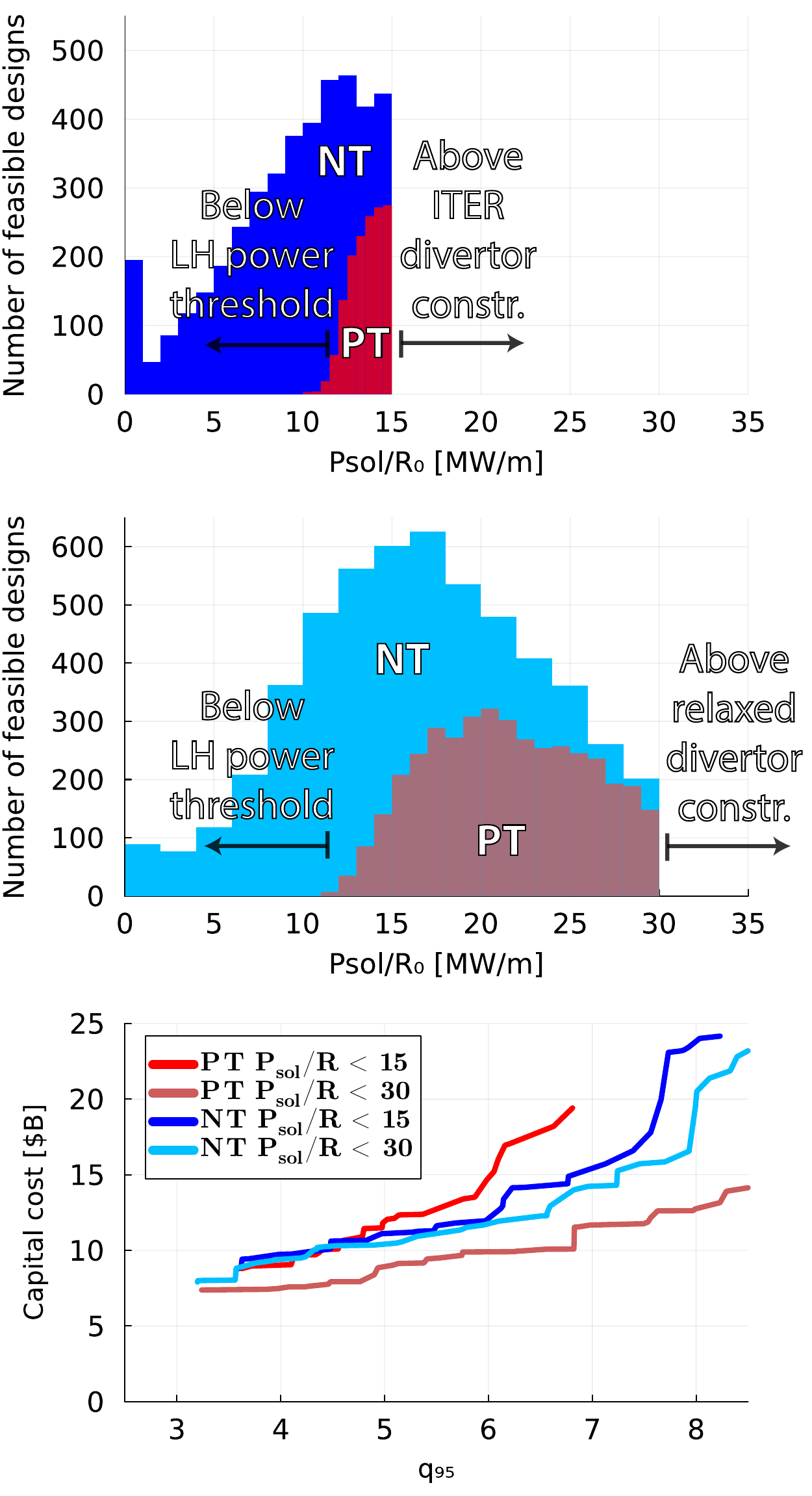}
    \caption{Power exhaust operational windows and their impact on power plant optimization. \textbf{Top}: Under the research-established divertor limit ($P_{\text{sol}}/R < 15$ MW/m), PT configurations (red) are squeezed into a narrow operational window between the L-H power threshold and heat flux constraints, as indicated by the arrows. NT (purple) operates freely across the full allowable range. \textbf{Middle}: Relaxing to an advanced divertor limit of 30 MW/m dramatically expands PT's accessible design space, enabling operation at higher power densities previously forbidden. \textbf{Bottom}: Pareto fronts show PT incurs a \$2-10B cost penalty under conservative divertor limits that largely vanishes with advanced heat-handling technology, while NT maintains consistent economics regardless of divertor assumptions.}
    
    \label{fig:psol_r_campaign}
\end{figure}

Power exhaust management represents one of the most critical challenges for power plant design, with different implications for PT and NT configurations. Figure~\ref{fig:psol_r_campaign} reveals how power exhaust constraints  shape the accessible design space. For PT configurations, the operational window is squeezed between two competing requirements: maintaining sufficient power across the separatrix to sustain H-mode ($P_{\text{sol}} > P_{LH}$) while respecting engineering limits on parallel heat flux ($P_{\text{sol}}/R < 15$ MW/m).

The top panel demonstrates that PT designs cluster near the lower bound of the operational window, with many designs operating with minimal margin above the L-H threshold. This narrow window becomes particularly constraining for high-performance designs that require elevated power to maintain strong pedestals. In contrast, NT configurations, free from L-H threshold requirements, utilize the full range of allowable power exhaust values, providing greater operational flexibility.

Relaxing the power exhaust limit to $P_{\text{sol}}/R < 30$ MW/m—potentially achievable with advanced divertor concepts or by running detached plasmas—dramatically impacts the optimization results (bottom panel). For PT, this relaxation shifts the Pareto front downwards by 20-50\%. NT configurations show more modest benefits, as their design optimization is less constrained by power exhaust limits. This differential sensitivity suggests that advances in power handling technology would preferentially benefit PT power plant designs.

\begin{figure*}
    \centering
    \includegraphics[width=0.99\textwidth]{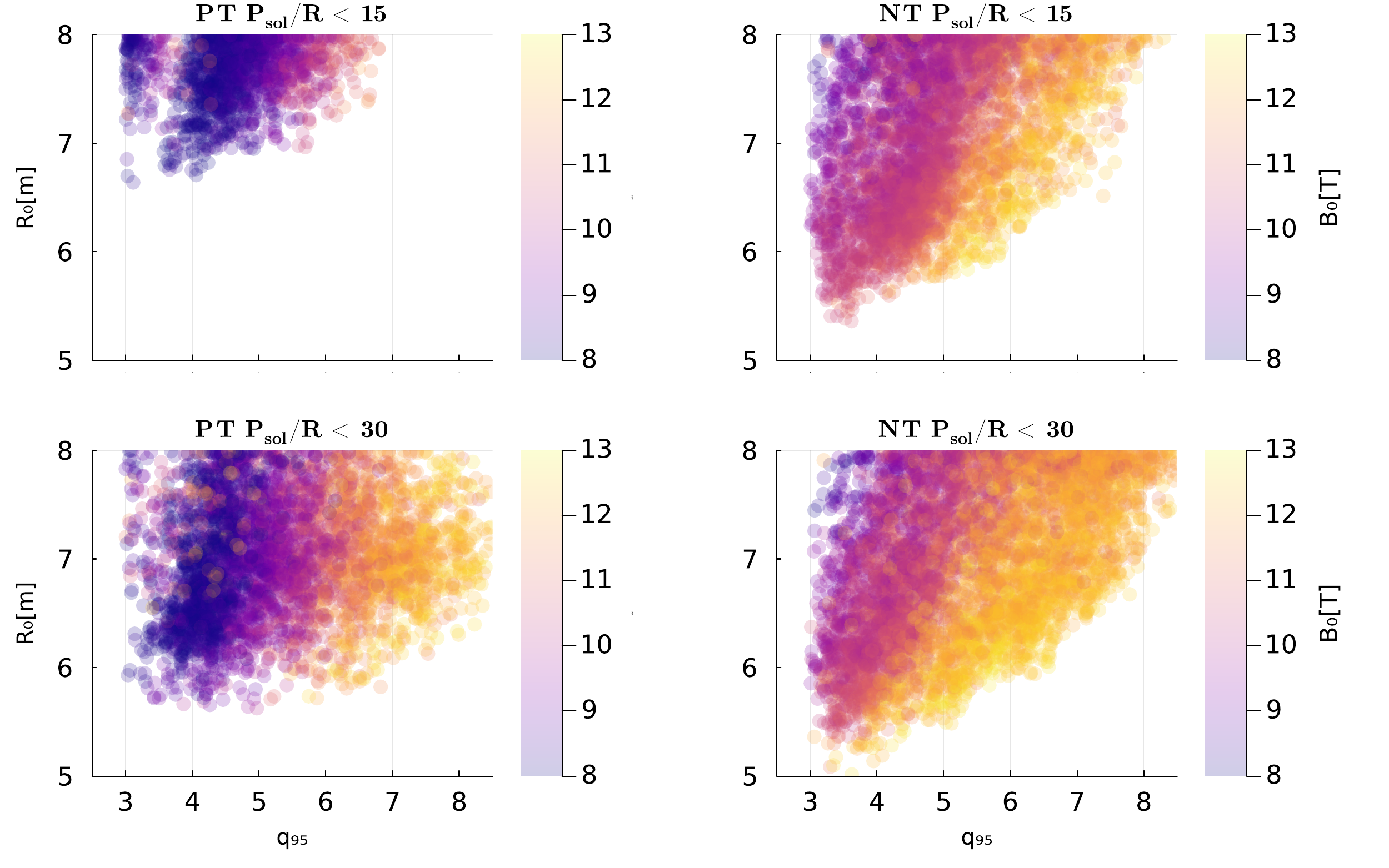}
    \caption{Parameter space evolution with relaxed power exhaust constraints. Under the strict $P_{\text{sol}}/R < 15$ MW/m limit (top), PT designs are forced toward larger major radii ($R_0 > 7$m) to distribute heat flux, while NT maintains access to more compact designs. Relaxing to 30 MW/m (bottom) enables PT to access previously inaccessible compact, high-field designs ($R_0 \sim 5.5$m, $B_0 > 12$T), substantially expanding its optimization space. NT shows only marginal expansion, confirming that power exhaust is not its primary design driver. This differential sensitivity explains the cost reductions shown in Figure~\ref{fig:psol_r_campaign}.}
    \label{fig:psol_r_campaign_4_scatter}
\end{figure*}

To better understand how power exhaust constraints shape the accessible parameter space, Figure~\ref{fig:psol_r_campaign_4_scatter} presents the feasible designs in terms of physical machine parameters rather than cost metrics. The major radius ($R_0$) versus safety factor ($q_{95}$) space reveals striking differences between configurations and constraint levels.

Under the stringent $P_{\text{sol}}/R < 15$ MW/m constraint (top panels), PT configurations cluster at larger major radii ($R_0 > 6.5$ m), unable to access the compact, high-field design space available to NT. This limitation arises from PT's need to maintain sufficient power crossing the separatrix for H-mode access while respecting divertor engineering limits. For PT designs, the constraining L-H threshold power scales as $P_{LH} \propto B_0^{0.8}$, forcing the optimizer to reduce magnetic field strength to maintain H-mode access within the available power window. NT configurations, free from L-H threshold requirements, populate the full range down to $R_0 \sim 5.5$ m, with many designs achieving $B_0 > 12$ T, a parameter space completely inaccessible to PT under these constraints.

Relaxing the power exhaust limit to 30 MW/m (bottom panels) reveals the potential in PT designs. The minimum achievable major radius drops to 5.7 m, with the newly accessible parameter space predominantly utilizing higher magnetic fields to maintain performance in more compact machines. This expansion explains the cost reductions noted in Table~\ref{tab:results_all_gen_combined}, as smaller machines inherently require less material. NT configurations show more modest gains, with the minimum radius decreasing only marginally to 5.5 m, confirming that power exhaust is not their primary limiting constraint. 

Our results provide quantitative support for aspects of both perspectives in the Federici-Creely debate. Consistent with Federici et al. \cite{federici2024relationship}, we find that TF coil costs dominate the economics of fusion power plants. However, by having more modest fusion power levels (800-1000 MW versus Federici's 2 GW), implementing bucked TF coil designs as advocated by Creely et al. \cite{creely2024comment}, and, critically, incorporating self-consistent radiative edge solutions that Federici's analysis did not consider, we identify viable compact designs that were previously dismissed.
The key insight is that Federici's power exhaust constraint (using a more complex variant of $P_{sol}/R$) creates an artificial barrier for PT configurations and does not consider radiative edge solutions. For example, our NT configurations operate with edge radiation and effectively circumvent the power exhaust limits that force Federici's designs to large size. Contrary to the ARC concept which assumes $P_{sol}/R > 50$ MW/m proposed by Creely et al., we chose not to explore $P_{sol}/R > 30$ MW/m as this is well beyond current experimental validation and potentially masks the fundamental importance of power exhaust constraints. 

\subsection{Impact of Physics Uncertainties}
The optimization results depend critically on the accuracy of physics models, particularly for edge transport and stability. We explore this sensitivity through systematic variations of key physics parameters.

\subsubsection{L-H Transition Threshold Uncertainty}

\begin{figure}[H]
    \centering
    \includegraphics[width=0.49\textwidth]{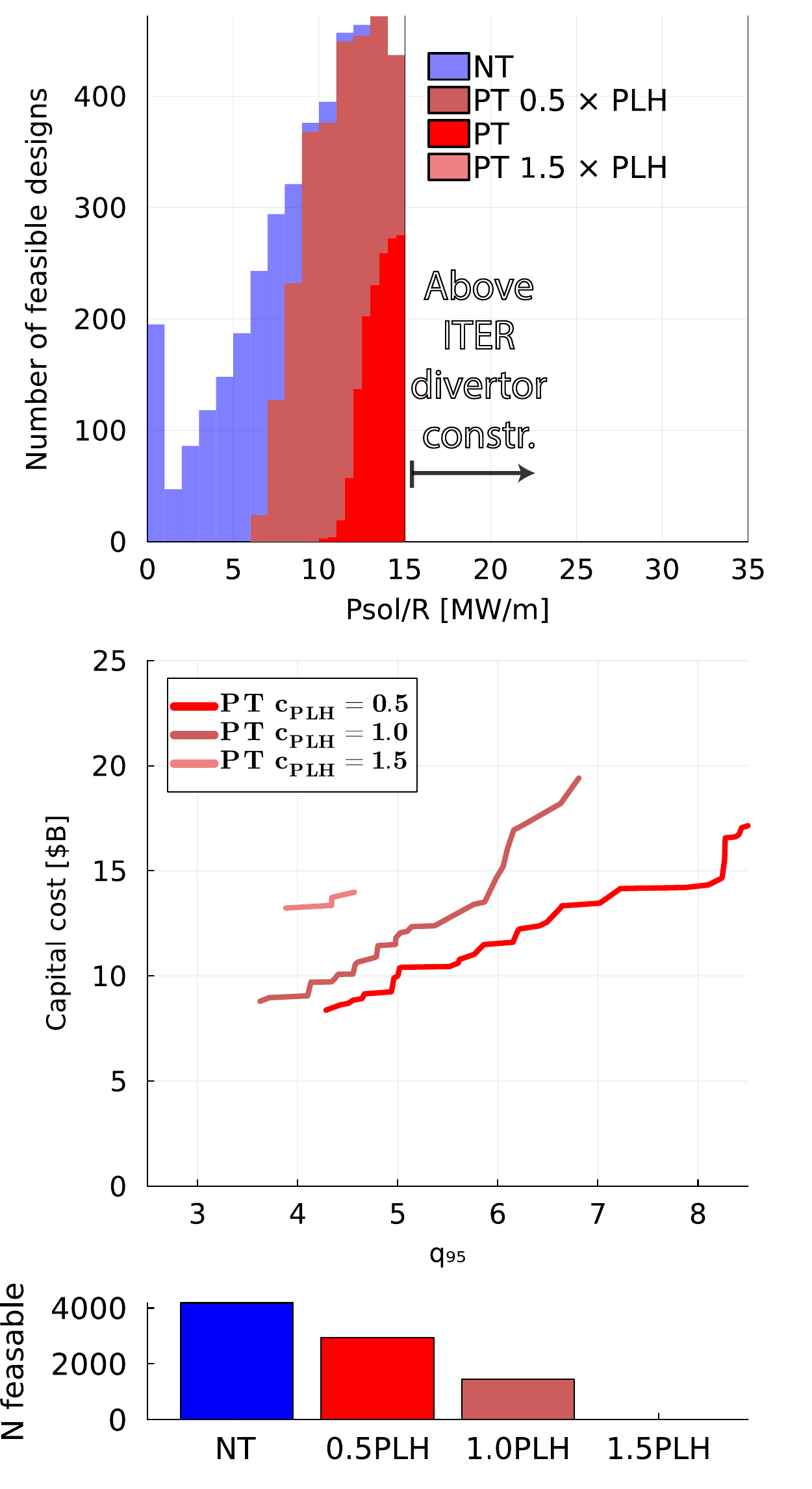}
    \caption{Sensitivity of PT design space to L-H power threshold uncertainty. \textbf{Top}: Distribution of feasible designs for NT and PT with varied L-H scaling factors (0.5×, 1.0×, 1.5×), note that for 1.5× the L-H scaling the design space completely vanishes. \textbf{Middle}: Pareto fronts showing cost implications of L-H threshold variations. \textbf{Bottom}: Total feasible designs demonstrating that 50\% uncertainty in L-H scaling translates to order-of-magnitude changes in viable PT configurations, while NT remains unaffected.}
    \label{fig:plh_campaign}
\end{figure}

The power threshold for L-H transition remains one of the largest uncertainties in tokamak physics, with experimental scaling laws showing factor-of-two variations depending on wall conditions, isotope mass, and configuration details. Figure~\ref{fig:plh_campaign} explores how this uncertainty propagates through the power plant design optimization.

Reducing the L-H threshold by 50\% ($c_{PLH} = 0.5$) dramatically expands the PT design space, increasing the number of feasible solutions by nearly an order of magnitude (bottom panel). This expanded operational space allows access to smaller, less expensive machines that were previously excluded by the H-mode access requirement. Conversely, if the L-H threshold proves higher than current scaling laws predict, the PT design space could become severely constrained, potentially eliminating the cost advantage over NT configurations. This sensitivity underscores the value of NT's inherent L-H independence for power plant design robustness.

\subsubsection{Pedestal Performance Degradation}

While the majority of modern tokamaks achieve their highest performance and confinement parameters in ELMy regimes, strong Type-I ELMs cannot be tolerated in a fusion power plant as the ELM energy fluence scales strongly with plasma current and reaches dangerous levels under fusion power plant conditions \cite{gunn2017surface}. The leading approach to mitigate this heat flux issue to operate fusion power plants under ELM-free or ELM-mitigated conditions. Numerous strategies for avoiding strong Type-I ELMs in PT power plants, including grassy ELMs, pellet pacing, RMP suppression, QH mode, QCE/EDA H-mode, I-mode and vertical kicks, are currently being explored for fusion power plant implementation \cite{viezzer2023prospects, paz-soldan2021plasma, nelson2021timedependent, kim2024highest}. However, these strategies may degrade pedestal performance below the ideal EPED predictions, as can variations in fueling expected in future machines \cite{viezzer2023prospects, paz-soldan2021plasma, kim2024highest, nelson_setting_2020}, which may in turn degrade core plasma performance.

\begin{figure}[H]
    \centering
    \includegraphics[width=0.49\textwidth]{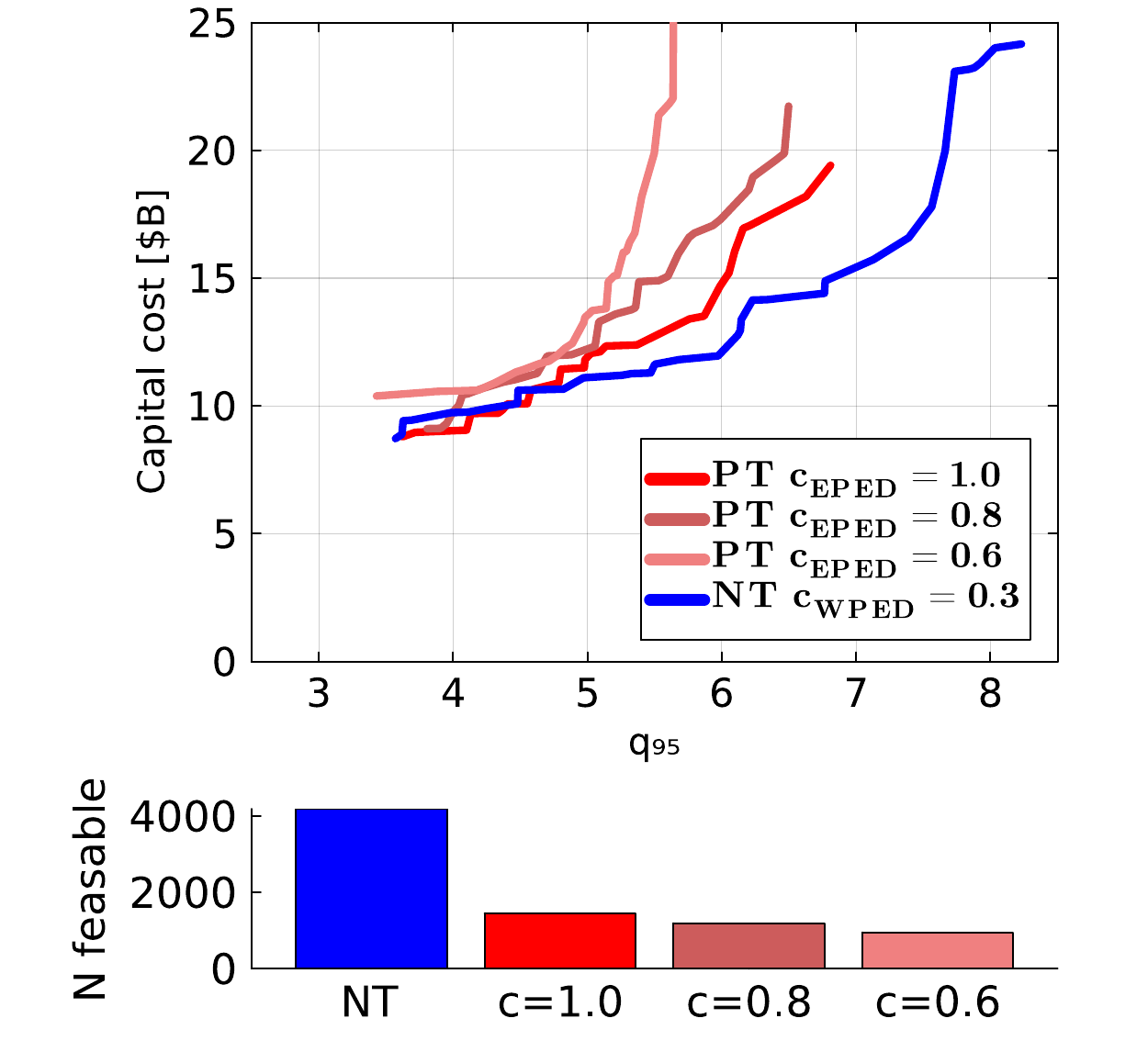}
    \caption{Impact of pedestal performance on power plant optimization. \textbf{Top}: Pareto fronts for PT with reduced EPED factors ($c_{\text{EPED}}$ = 1.0, 0.8, 0.6) representing ELM mitigation scenarios, compared to NT with nominal WPED model. \textbf{Bottom}: Total feasible designs showing PT sensitivity to pedestal degradation versus the robust NT design space that operates without ELMs.}
    \label{fig:eped_campaign}
\end{figure}

Figure~\ref{fig:eped_campaign} offers a strategy to characterize the potential impact of ELM mitigation on fusion power plant performance by reducing the pedestal multiplication factor from the ideal $c_{\text{EPED}} = 1.0$ to degraded values of 0.8 and 0.6, representing 20\% and 40\% pedestal degradation respectively. The Pareto fronts reveal remarkable insensitivity to moderate pedestal degradation: the cost-performance trade-off remains nearly identical for $c_{\text{EPED}}$ values from 1.0 to 0.6 up to $q_{95}=5$. This robustness arises from the optimizer's ability to compensate for reduced pedestal pressure through increased plasma current (reaching 20.8 MA for $c_{\text{EPED}} = 0.6$) and adjusted profiles. However, this compensation comes at the expense of increased current drive power requirements and limits operation at higher $q_{95}$ for the 40\% reduced pedestal study.

The bottom panel reveals a key insight into what limits PT power plant design: the number of feasible designs decreases with pedestal degradation, but not as dramatically as one might expect from pure confinement considerations. This modest sensitivity occurs because the PT design space is not primarily performance-limited but rather constrained by the narrow operational window between the L-H threshold requirement ($P_{\text{sol}} > P_{LH}$) and the power exhaust engineering limit ($P_{\text{sol}}/R < 15$ MW/m). 

With degraded pedestals, designs must operate at higher current to maintain fusion power output, which increases the power crossing the separatrix. This pushes more designs against the $P_{\text{sol}}/R$ limit while still needing to exceed $P_{LH}$. The fact that significant feasible solutions remain even with 40\% pedestal degradation demonstrates that power exhaust constraints, not core confinement quality, represent the primary bottleneck for PT power plant optimization. This finding reinforces the critical importance of advanced divertor solutions for enabling PT power plant designs, regardless of pedestal performance achievements. Sensitivity analysis of the WPED model parameter was performed on the NT designs by varying $c_{WPED}$ from its nominal value of 0.3. Optimization with $c_{WPED} = 0.2$ produced nearly identical Pareto fronts to the baseline case, with only marginal differences in optimal machine parameters. This insensitivity contrasts sharply with the strong dependence observed for PT configurations on $c_{EPED}$ variations, likely due to the extended transport boundary at $\rho = 0.9$ for NT reducing the relative importance of the edge energy fraction.

\subsection{Technology Impact: Superconductor Comparison} \label{sec:HTS_LTS} 
\begin{figure}[ht]
    \centering
    \includegraphics[width=0.49\textwidth]{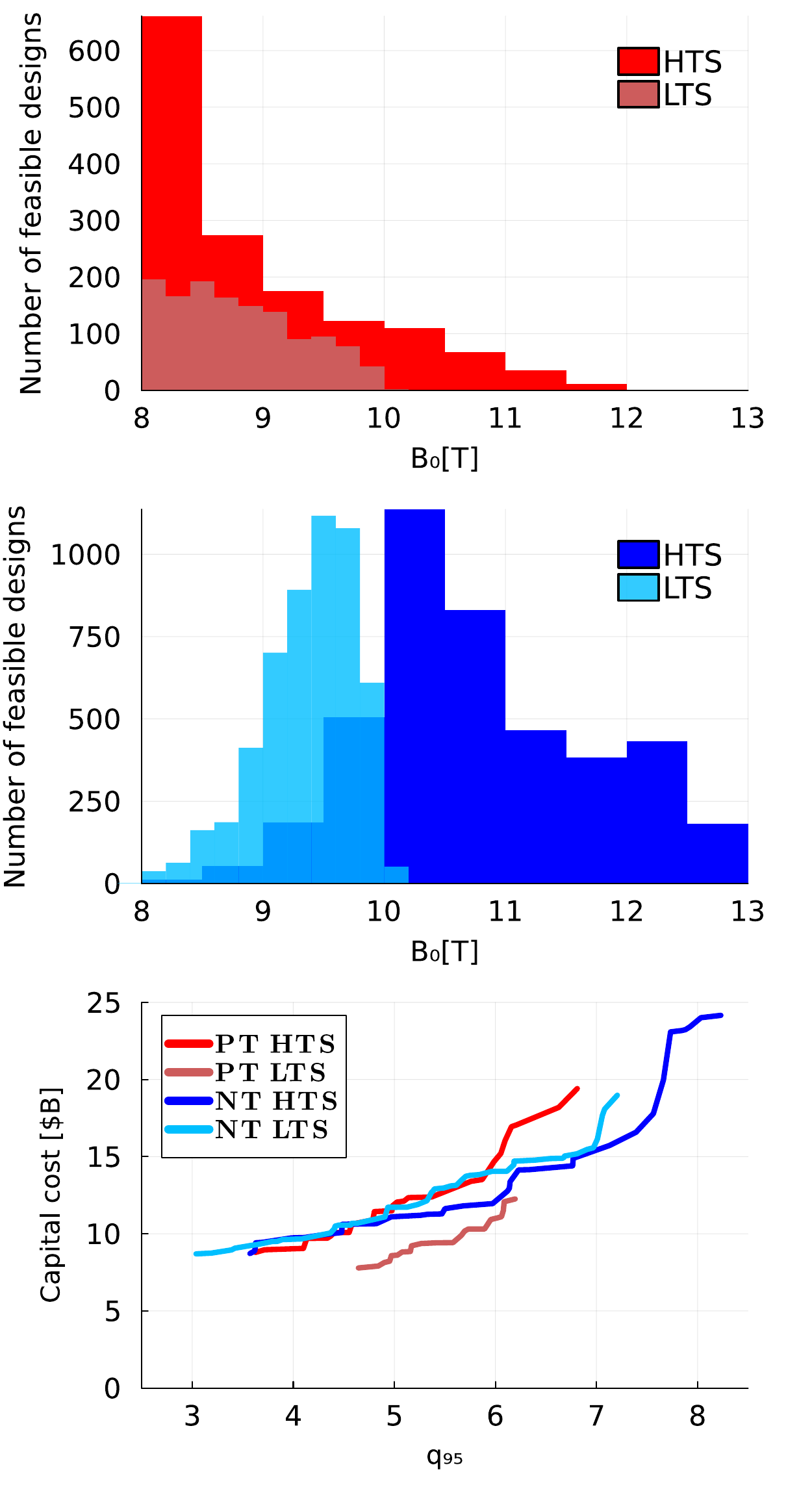}
    \caption{Comparison of high-temperature (HTS) and low-temperature (LTS) superconductor technologies. \textbf{Top} PT and \textbf{Middle} NT Magnetic field distributions showing HTS enables higher fields. \textbf{Bottom}: Pareto fronts revealing that for PT configurations, the $P_{\text{sol}}/R$ constraint negates the potential advantages of HTS technology, while NT benefits from the expanded operational space at higher fields.}
    \label{fig:hts_lts_campaign}
\end{figure}

The choice between low-temperature (LTS) and high-temperature (HTS) superconductor technology affects the accessible design space through magnetic field limitations. Figure~\ref{fig:hts_lts_campaign} compares optimization results for both technologies, revealing configuration-dependent benefits.

The economic analysis assumes future superconductor costs reflecting anticipated scale-up from fusion deployment. For ReBCO HTS, we set \$67/kA-m (equivalent to \$7,000/kg), representing a reduction from current levels of \$150-200/kA-m toward the \$50/kA-m threshold for widespread adoption \cite{nextbigfuture2023}. This projection aligns with demand-driven cost reductions, particularly from projects like Commonwealth Fusion's SPARC \cite{creely2020overview} requiring 10,000 km of superconducting wire. For N$b_3$Sn LTS, we use \$700/kg based on historical accelerator magnet costs \cite{uspas2018}, though current pricing likely exceeds \$1,000/kg. These assumptions reflect nth-of-a-kind economics where manufacturing has achieved economies of scale.

HTS technology enables operation at higher magnetic fields (up to 15 T on-axis versus ~10 T for LTS), which should allow more compact, less expensive designs. However, the predicted benefits depend strongly on the configuration. For PT designs, the constraining L-H threshold power is proportional to the field strength:  $P_{LH} \propto B_0^{0.8}$ and thus prevents full utilization of higher field capability. The power exhaust constraint prevents the optimizer from reducing machine size without violating the $Psol/R$ limit. Consequently, HTS provides only marginal benefits for PT configurations. 

NT configurations, with their broader power exhaust windows and higher auxiliary power requirements, better exploit HTS capabilities. The ability to operate at higher field partially compensates for the lack of pedestal pressure, enabling NT-HTS designs to approach cost equivalence with PT-LTS configurations. This suggests that HTS technology development may be particularly valuable for advancing NT power plant designs. These optimization results warrant careful interpretation in light of engineering realities not fully captured in the current modeling. The optimizer readily exploits the diminishing returns of critical current density versus magnetic field for LTS technology \cite{lu2009field}, pushing designs to maximum fields approaching 18 T with proportionally larger magnet cross-sections to compensate for reduced current density. While we implemented a 10\% current margin in our analysis, practical LTS magnets require additional engineering temperature margin, typically operating at 5.2 K rather than the theoretical 4.2 K limit, which substantially degrades superconductor performance. Furthermore, detailed thermal-hydraulic analysis by Wen et al. \cite{wen2023thermal} demonstrated that nuclear heating causes approximately 1 K temperature rise at distances of ~160 meters from the helium inlet in CFETR TF coils, adding another layer of thermal margin erosion not captured in our optimization. This suggests our LTS field predictions may be optimistic by 18\%, estimated using the scaling laws given in \cite{lu2009field}, and highlights the need for future work coupling neutronics calculations with detailed TF cooling design. The apparent 30\% cost advantage of LTS over HTS for PT configurations may therefore be eroded when accounting for these engineering realities and the challenges of operating near performance limits.
Looking forward, technology cost trends may further shift the economic balance. While our analysis assumes mature manufacturing costs for both technologies, the future cost may even be lower due to the rapidly expanding HTS supply chain driven by fusion demonstration projects. Conversely, the LTS supply chain faces potential constraints as demand shifts toward HTS, potentially reversing the current cost differential. These supply chain dynamics, while speculative, reinforce the strategic value of maintaining design flexibility for both superconductor technologies as the fusion industry matures.

\subsection{Balance of Plant Thermal Cycle Optimization}
The optimization framework allowed free selection between feedwater Rankine and Brayton thermal cycles to maximize plant efficiency across the 800-1200 MW thermal power range typical of our designs. Both cycles demonstrated viable operation through self-consistent optimization of inlet/outlet temperatures and mass flow rates, achieving overall balance of plant efficiencies ($\eta_{BOP} = P_{thermal}/P_{electric}$) between 34\% and 41\%.
Brayton cycles showed a modest advantage of several percentage points in thermal efficiency compared to Rankine cycles across the design space. However, this theoretical performance benefit must be weighed against the substantially higher complexity of Brayton cycle implementation, particularly for the gas handling and high-temperature heat exchanger systems.
The systematic difference in thermal efficiency between NT and PT configurations in Table A-1 arises from their distinct operating regimes. NT configurations, requiring approximately 1100 MW thermal power to achieve 200-300MW net electric, operate in a less favorable efficiency regime compared to PT designs that achieve the same electrical output with only 850 MW thermal power. This efficiency penalty at higher thermal powers partially offsets NT's other advantages.
Given the modest efficiency gains and the mature industrial base for Rankine cycle components in this power range, practical power plant designs may favor the proven reliability and reduced complexity of feedwater Rankine systems over the marginally more efficient Brayton alternative. This engineering judgment, while not enforced in our optimization, aligns with conventional power industry practice for facilities in the 1 GW thermal class.

\subsection{Assessment of Edge Stability}As discussed above, one of the largest differences between the NT and PT cases is the edge boundary condition that is set by pedestal physics. For NT plasmas in particular, it is well demonstrated that the upper limit for the pressure gradient in the edge region is set by ideal (infinite-$n$) ballooning modes \cite{nelson_robust_2023, nelson_first_2024, nelson_prospects_2022}. However, it is also evident from the DIII-D datasets that this maximum possible gradient is not always reached, with the physics understandings of the mechanisms responsible for setting the height and width of the ELM-free pedestal still not advanced enough for a fully predictive model such as that provided by EPED for PT plasmas \cite{nelson_characterization_2024}. Indeed, this uncertainty is the main motivation for the adoption of the WPED model described in Section~\ref{sec:EXP_WPED}, which aims to capture the characteristic behavior of the ELM-free NT pedestal on DIII-D for inclusion into the FUSE CMOOP framework. 

\begin{figure*}[ht]
    \centering
    \labelphantom{fig:baloo-a}
    \labelphantom{fig:baloo-b}
    \labelphantom{fig:baloo-c}
    \includegraphics[width=1\textwidth]{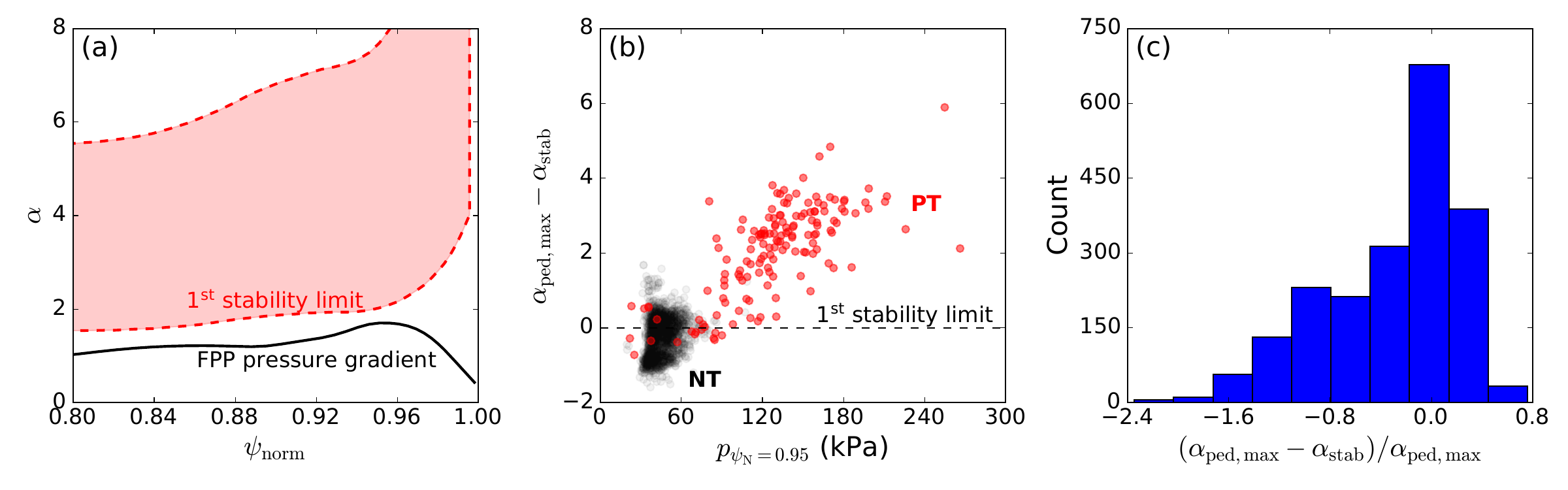}
    \caption{\textbf{Left}: Ideal ballooning stability limits (red) and equilibrium pressure gradient (black) shown for a representative NT solution. \textbf{Center}: The distance to the first stability limit for the full solution spaces of NT plasmas with $c_\mathrm{WPED}=0.3$ (black) and PT plasmas with $c_\mathrm{EPED}=1$ (red) shown as a function of the pedestal pressure. \textbf{Right}: Histogram of the distribution of NT solutions as a function of their normalized distance to the ideal stability limit.}
    \label{fig:baloo}
\end{figure*}

While ideal ballooning stability calculations are not yet directly included in the FUSE optimizer, it is possible to still assess the performance of this approach by running a code like BALOO \cite{snyder_edge_2002} over the solution space for a particular CMOOP execution. For a particular representative solution, this is demonstrated in Figure~\ref{fig:baloo}, which shows the normalized pressure gradient $\alpha$ as a function of the plasma normalized radius $\psi_\mathrm{N}$ in the edge region. For the entire pedestal region, this equilibrium $\alpha$ remains below the ideal stability limit for ballooning modes (shown in red). Further, for this particular solution (which has $\delta=-0.47$ and $c_\mathrm{WPED}=0.3$) and for the rest of the NT solution space, access to the second stability region for ideal ballooning modes is closed, demonstrating consistency between the FUSE solutions and the established physical understanding for the prevention of ELMy behavior at strong negative triangularity \cite{nelson_prospects_2022}.

To demonstrate this behavior for the full CMOOP solution space, we plot the distance between the peak of the normalized pressure gradient and the first stability limit for entire NT (with $c_\mathrm{WPED}=0.3$) and PT ($c_\mathrm{EPED}=1$) datasets as a function of the equilibrium pressure at $\psi_\mathrm{N}=0.95$ in Figure~\ref{fig:baloo}. As expected, the entire NT dataset lies near or below the 1$^\mathrm{st}$ stability limit, much in the same fashion as was observed in a systematic study of the experimental ELM-free dataset on DIII-D \cite{nelson_first_2024}. The distribution of NT cases in this solution space as a function of their stability metric is shown in Figure~\ref{fig:baloo}. The consistency between the bulk of the solution space and the expected ballooning stability metric suggests that the use of the $c_\mathrm{WPED}=0.3$ constraint for NT plasmas is a reasonable procedure for the study of the NT fusion power plant operational space, at least until a more robust predictive model (such as that provided by EPED for PT plasmas) is developed by the community.

\newpage

\section{Conclusions and Future Work}
\label{sec:conclusion}

This comprehensive optimization study reveals that the choice between positive and negative triangularity configurations for fusion power plants hinges not on achieving specific plasma performance metrics, but rather on navigating different operational constraints and their associated uncertainties.

Our analysis challenges several preconceptions about both configurations. PT designs demonstrate remarkable resilience to pedestal degradation, with the optimizer successfully compensating for up to 40\% reduction in pedestal height through increased plasma current and profile adjustments. This robustness in core performance, however, masks a more fundamental vulnerability: PT configurations operate within an increasingly narrow corridor between the L-H power threshold required for H-mode access and the power exhaust limits imposed by divertor technology. This operational window, rather than pedestal quality, emerges as the primary constraint shaping PT power plant designs.

The sensitivity analyses reveal how uncertainties in extrapolating current physics understanding to power plant scales could dramatically alter the competitive landscape. A 50\% variation in L-H threshold scaling—well within current experimental uncertainties—can either eliminate viable PT designs or substantially expand their parameter space. Similarly, advancing divertor technology to handle 30 MW/m rather than the research-established benchmark of 15 MW/m would enable PT access to compact, high-field designs currently exclusive to NT, reducing capital costs by 15-20\%.

NT configurations offer a different value proposition. While requiring approximately 20\% higher baseline capital investment, they operate in an unconstrained regime free from L-H threshold requirements and with substantially relaxed power exhaust limits. With these relaxed constraints, NT configurations provide more design space options for commercial fusion power plants.

The optimization revealed unexpected technology synergies that could reshape future development priorities. NT's natural evolution toward compact, high-field designs ($R_0$ $\sim$ 5.5 m, $B_0$ $>$ 12 T) creates strong synergy with high-temperature superconductor development. In contrast, PT's optimization toward larger, low-field configurations ($R_0$ $>$ 6.5 m, $B_0$ $\sim$ 8 T) derives limited benefit from HTS technology due to power exhaust constraints. This suggests that continued HTS maturation could preferentially advance NT pathways, potentially offsetting their higher baseline costs.

These findings argue for a risk-informed approach to fusion development that considers both economic optimization and operational uncertainties. For near-term demonstration plants where operational parameters can be carefully controlled and conservative assumptions verified, PT configurations offer the lowest capital cost path to net electricity generation. However, for commercial power plants that must operate reliably across diverse conditions with significant uncertainties in physics extrapolation, NT's broader operational space and reduced sensitivity to constraint uncertainties may justify its cost premium.

The systematic exploration of over 200,000 integrated designs demonstrates the value of high-fidelity optimization in revealing non-intuitive design drivers. Power exhaust management, rather than core confinement quality, emerges as the primary differentiator between configurations, highlighting the critical importance of advancing divertor technology, as well as improving our predictive capability for plasma-material interactions. These findings quantitatively inform the Federici-Creely debate on achievable power plant designs, providing insight on their opposing views by demonstrating how radiative edge solutions and integrated optimization enable more compact configurations within realistic power exhaust constraints. As the fusion community progresses toward power plant design, the framework and insights presented here provide a foundation for configuration selection driven by quantitative trade-offs, not historical inertia or speculative reasoning.

Building on the recent techno-economic study by Ward et al.,\cite{ward2025fusionhighvalueheatproduction}, future work will extend our optimization studies to value process-heat sales alongside electricity. We plan to introduce a “HeatRevenueActor” that links blanket outlet temperature, total thermal output and recirculating-power fraction to a levelised cost-of-heat objective, enabling FUSE to search for blanket and balance of plant designs that maximize high-value heat while retaining acceptable physics and engineering margins.

\ack
The authors thank G. Avdeeva and N. Shi for their participation in team discussions during the project development. We also thank C. Akcay, G. Avdeeva and H. S. Wilson for reviewing the manuscript. These results are made possible by support from the U.S. Department
of Energy under the SMARTs grant DE-SC0024399 and DE-SC0022270 as well as General Atomics corporate funding. This material is based upon work supported by the U.S. Department of Energy, Office of Science, Office of Fusion Energy Sciences, using the DIII-D National Fusion Facility, a DOE Office of Science user facility, under Award DE-FC02-04ER54698.


\section*{Disclaimer}
This report was prepared as an account of work sponsored by an agency of the United States Government.
Neither the United States Government nor any agency thereof, nor any of their employees, makes any
warranty, express or implied, or assumes any legal liability or responsibility for the accuracy, completeness,
or usefulness of any information, apparatus, product, or process disclosed, or represents that its use would not
infringe privately owned rights. Reference herein to any specific commercial product, process, or service by
trade name, trademark, manufacturer, or otherwise, does not necessarily constitute or imply its endorsement,
recommendation, or favoring by the United States Government or any agency thereof. The views and opinions
of authors expressed herein do not necessarily state or reflect those of the United States Government or any
agency thereof.

\section*{References}
\bibliographystyle{iopart-num}
\bibliography{2025FUSENT}

\appendix
\renewcommand{\thetable}{A-\arabic{table}}
\setcounter{table}{0}
\section{Design-Space Statistics}
\label{apendix:Design_space_statistics}
\begin{sidewaystable*}[ht]
    \vspace{0.5\textwidth}

  \centering
  \caption{Statistical summary of feasible fusion power plant designs across all 50 generations of optimization, and separately for generations 40--50 (converged portion). Values represent mean ± standard deviation for key design parameters and performance metrics across the population of designs satisfying all constraints.}
  \label{tab:results_all_gen_combined}
  \resizebox{\textheight}{!}{%
  \begin{tabular}{lcccccccccccccc}
    \hline
    \textbf{Study} & \textbf{$R_0$ [m]} & \textbf{$I_p$ [MA]} & \textbf{$B_0$ [T]}
      & \textbf{$\delta$} & \textbf{$\kappa$} & \textbf{$z_{eff,ped}$}
      & \textbf{$f_{GW}$} & \textbf{$P_{aux}$ [MW]}
      & \textbf{Capital Cost [\$B]} & \textbf{$q_{95}$}
      & \textbf{$P_{electric,net}$ [MW]} & \textbf{$\eta_{thermal}$ [\%]}
      & \textbf{$P_{sol}$ [MW]} & \textbf{$P_{rad,tot}$ [MW]}\\
    \hline
    \hline
    PT                                             & $7.6 \pm 0.3$ & $18.5 \pm 1.3$ & $-8.9 \pm 0.9$ & $0.6 \pm 0.1$ & $1.8$ & $1.7 \pm 0.1$ & $0.7 \pm 0.1$ & $21.8 \pm 9.7$ & $14.7 \pm 3.6$ & $4.7 \pm 0.8$ & $289.8 \pm 29.9$ & $39.5 \pm 1.8$ & $103.5 \pm 8.2$  & $-93.1 \pm 16.4$ \\
    NT                                             & $7.0 \pm 0.7$ & $16.0 \pm 2.2$ & $-10.8 \pm 0.9$& $-0.4 \pm 0.0$& $1.8$& 1.5            & $1.1 \pm 0.2$ & $48.3 \pm 9.1$ & $18.1 \pm 7.9$ & $5.0 \pm 1.1$  & $344.2 \pm 69.6$& $35.7 \pm 1.8$ & $66.9 \pm 28.8$ & $-211.7 \pm 47.3$ \\
    PT (\(P_{\text{sol}}/R<30\))                   & $7.0 \pm 0.5$ & $17.2 \pm 1.8$ & $-10.0 \pm 1.5$& $0.7 \pm 0.1$ & $1.8$& $1.6 \pm 0.1$ & $0.8 \pm 0.1$ & $38.4 \pm 15.1$& $14.2 \pm 4.0$ & $5.4 \pm 1.3$ & $359.7 \pm 86.5$ & $38.1 \pm 1.6$ & $150.1 \pm 31.0$& $-117.5 \pm 31.6$ \\
    NT (\(P_{\text{sol}}/R<30\))                   & $6.9 \pm 0.7$ & $15.5 \pm 2.3$ & $-11.2 \pm 1.0$& $-0.4 \pm 0.0$& $1.8$& 1.5            & $1.1 \pm 0.1$ & $55.5 \pm 12.6$& $19.0 \pm 8.4$ & $5.4 \pm 1.3$ & $404.8 \pm 100.5$& $35.0 \pm 2.0$ & $115.7 \pm 47.5$& $-219.7 \pm 46.4$ \\
    PT (\(c_{\text{PLH}}=0.5\))                    & $7.6 \pm 0.3$ & $17.4 \pm 2.0$ & $-10.3 \pm 1.4$& $0.6 \pm 0.2$ & $1.8$& $1.6 \pm 0.1$ & $0.9 \pm 0.1$ & $22.2 \pm 10.1$& $16.6 \pm 4.7$ & $5.7 \pm 1.4$ & $305.1 \pm 48.2$ & $38.8 \pm 2.1$ & $87.6 \pm 16.0$ & $-119.3 \pm 26.3$ \\
    PT (\(c_{\text{EPED}}=0.8\))                   & $7.7 \pm 0.3$ & $19.6 \pm 1.2$ & $-9.6 \pm 1.0$ & $0.6 \pm 0.0$ & $1.8$& $1.7 \pm 0.1$ & $0.7 \pm 0.1$ & $32.8 \pm 9.1$ & $16.5 \pm 4.4$ & $4.8 \pm 0.8$ & $284.9 \pm 29.5$ & $39.1 \pm 1.7$ & $107.0 \pm 7.0$ & $-103.7 \pm 16.4$ \\
    PT (\(c_{\text{EPED}}=0.6\))                   & $7.6 \pm 0.3$ & $20.8 \pm 0.7$ & $-10.0 \pm 0.8$& $0.6 \pm 0.0$ & $1.8$& $1.6 \pm 0.1$ & $0.7 \pm 0.1$ & $38.9 \pm 7.0$ & $16.1 \pm 4.1$ & $4.6 \pm 0.5$ & $275.0 \pm 19.9$ & $39.0 \pm 1.4$ & $108.9 \pm 5.1$ & $-106.6 \pm 13.0$ \\
    PT (LTS)                                       & $7.7 \pm 0.2$ & $18.3 \pm 1.3$ & $-8.8 \pm 0.5$ & $0.6 \pm 0.1$ & $1.8$& $1.7 \pm 0.1$ & $0.7 \pm 0.1$ & $24.7 \pm 8.4$ & $11.5 \pm 2.4$ & $4.8 \pm 0.7$ & $283.9 \pm 24.3$ & $39.8 \pm 1.7$ & $104.5 \pm 8.1$  & $-90.6 \pm 14.3$ \\
    NT (LTS)                                       & $7.4 \pm 0.4$ & $15.9 \pm 1.8$ & $-9.4 \pm 0.4$ & $-0.4 \pm 0.0$& $1.8$& 1.5            & $1.2 \pm 0.1$ & $59.4 \pm 9.6$ & $14.5 \pm 6.0$ & $4.6 \pm 0.8$ & $349.5 \pm 71.9$ & $35.1 \pm 1.8$ & $63.2 \pm 30.4$ & $-241.9 \pm 46.1$ \\
    \hline
    \multicolumn{15}{l}{\textbf{Subset: Generations 40--50 (Converged Portion)}} \\
    \hline
    \hline
    PT                                             & $7.5 \pm 0.3$ & $17.8 \pm 1.2$ & $-8.9 \pm 0.9$ & $0.7 \pm 0.0$ & $1.8$& $1.7 \pm 0.1$ & $0.8 \pm 0.1$ & $21.2 \pm 9.2$ & $14.0 \pm 3.4$ & $4.8 \pm 0.8$ & $289.6 \pm 28.8$ & $39.6 \pm 1.9$ & $102.4 \pm 8.0$  & $-92.0 \pm 16.6$ \\
    NT                                             & $6.8 \pm 0.7$ & $14.4 \pm 1.5$ & $-11.4 \pm 0.9$& $-0.4 \pm 0.0$& $1.8$& 1.5            & $1.1 \pm 0.1$ & $49.3 \pm 8.6$ & $17.3 \pm 6.9$ & $5.6 \pm 1.2$ & $330.9 \pm 62.6$ & $35.7 \pm 1.7$ & $65.5 \pm 28.0$ & $-205.8 \pm 47.3$ \\
    PT (\(P_{\text{sol}}/R<30\))                   & $6.6 \pm 0.5$ & $15.8 \pm 1.5$ & $-10.5 \pm 1.7$& $0.7 \pm 0.0$ & $1.8$& $1.6 \pm 0.1$ & $0.8 \pm 0.1$ & $39.5 \pm 14.6$& $12.8 \pm 3.4$ & $6.0 \pm 1.5$ & $336.8 \pm 72.5$ & $38.3 \pm 1.4$ & $145.1 \pm 27.9$& $-110.0 \pm 26.3$ \\
    NT (\(P_{\text{sol}}/R<30\))                   & $6.7 \pm 0.8$ & $14.2 \pm 2.0$ & $-11.8 \pm 0.7$& $-0.4 \pm 0.0$& $1.8$& 1.5            & $1.2 \pm 0.1$ & $55.0 \pm 10.9$& $17.2 \pm 6.3$ & $5.9 \pm 1.4$ & $398.4 \pm 100.9$& $35.2 \pm 1.7$ & $112.5 \pm 45.1$& $-214.0 \pm 44.0$ \\
    PT (\(c_{\text{PLH}}=0.5\))                    & $7.4 \pm 0.4$ & $16.0 \pm 1.1$ & $-10.1 \pm 1.6$& $0.7 \pm 0.1$ & $1.8$& $1.6 \pm 0.1$ & $0.9 \pm 0.1$ & $18.4 \pm 9.0$ & $15.1 \pm 4.1$ & $6.3 \pm 1.5$ & $295.5 \pm 35.9$ & $39.5 \pm 2.1$ & $83.9 \pm 15.8$ & $-108.6 \pm 18.6$ \\
    PT (\(c_{\text{EPED}}=0.8\))                   & $7.6 \pm 0.3$ & $19.2 \pm 1.1$ & $-9.6 \pm 1.0$ & $0.7 \pm 0.0$ & $1.8$& $1.7 \pm 0.1$ & $0.7 \pm 0.1$ & $32.6 \pm 9.0$ & $16.1 \pm 4.8$ & $4.9 \pm 0.8$ & $282.9 \pm 26.6$ & $39.3 \pm 1.8$ & $106.3 \pm 7.1$ & $-102.0 \pm 16.1$ \\
    PT (\(c_{\text{EPED}}=0.6\))                   & $7.6 \pm 0.3$ & $20.7 \pm 0.7$ & $-10.1 \pm 0.8$& $0.6 \pm 0.0$ & $1.8$& $1.6 \pm 0.1$ & $0.7 \pm 0.1$ & $37.9 \pm 5.9$ & $15.9 \pm 4.2$ & $4.7 \pm 0.5$ & $273.2 \pm 16.6$ & $39.3 \pm 1.4$ & $107.6 \pm 4.9$ & $-103.9 \pm 10.8$ \\
    PT (LTS)                                       & $7.7 \pm 0.2$ & $17.7 \pm 1.1$ & $-8.8 \pm 0.5$ & $0.7 \pm 0.0$ & $1.8$& $1.7 \pm 0.1$ & $0.8 \pm 0.1$ & $24.4 \pm 8.1$ & $11.0 \pm 2.2$ & $4.9 \pm 0.7$ & $286.1 \pm 25.4$ & $39.7 \pm 1.7$ & $104.5 \pm 7.4$  & $-91.1 \pm 14.0$ \\
    NT (LTS)                                       & $7.3 \pm 0.5$ & $14.9 \pm 1.5$ & $-9.5 \pm 0.3$ & $-0.4 \pm 0.0$& $1.8$& 1.5            & $1.3 \pm 0.1$ & $62.6 \pm 8.4$ & $14.2 \pm 5.3$ & $4.9 \pm 0.8$ & $336.2 \pm 62.7$ & $35.2 \pm 1.8$ & $62.2 \pm 30.6$ & $-239.9 \pm 43.2$ \\
    \hline
  \end{tabular}}
\end{sidewaystable*}

\end{document}